\documentclass[10pt,letterpaper]{article}
\usepackage[top=0.85in,left=2.75in,footskip=0.75in]{geometry}

\usepackage{amsmath,amssymb}

\usepackage{changepage}

\usepackage[utf8x]{inputenc}

\usepackage{textcomp,marvosym}

\usepackage{cite}

\usepackage{nameref,hyperref}

\usepackage[right]{lineno}

\usepackage{microtype}
\DisableLigatures[f]{encoding = *, family = * }

\usepackage[table]{xcolor}

\usepackage{array}

\newcolumntype{+}{!{\vrule width 2pt}}

\newlength\savedwidth

\raggedright
\usepackage[aboveskip=1pt,labelfont=bf,labelsep=period,justification=raggedright,singlelinecheck=off]{caption}

\makeatletter
\renewcommand{\@biblabel}[1]{\quad#1.}
\makeatother

\usepackage{lastpage,fancyhdr,graphicx}
\usepackage{epstopdf}
\fancyheadoffset[L]{2.25in}
\fancyfootoffset[L]{2.25in}
\begin{document}
\vspace*{0.2in}

\begin{flushleft}
{\Large
\textbf\newline{Understanding Political Communication and Political Communicators on Twitch} 
}
\newline
\\
Sangyeon Kim\textsuperscript{1,2*}
\\
\bigskip
\textbf{1} Observatory on Social Media, Indiana University, Bloomington, IN, USA
\\
\textbf{2} Luddy School of Informatics, Computing, and Engineering, Indiana University, Bloomington, IN, USA
\\
\bigskip

* ski15@iu.edu

\end{flushleft}
\section*{Abstract}
As new technologies rapidly reshape patterns of political communication, platforms like Twitch are transforming how people consume political information. This entertainment-oriented live streaming platform allows us to observe the impact of technologies such as ``live-streaming'' and ``streaming-chat'' on political communication. Despite its entertainment focus, Twitch hosts a variety of political actors, including politicians and pundits. This study explores Twitch politics by addressing three main questions: 1) Who are the political Twitch streamers? 2) What content is covered in political streams? 3) How do audiences of political streams interact with each other? To identify political streamers, I leveraged the Twitch API and supervised machine-learning techniques, identifying 574 political streamers. I used topic modeling to analyze the content of political streams, revealing seven broad categories of political topics and a unique pattern of communication involving context-specific ``emotes.'' Additionally, I created user-reference networks to examine interaction patterns, finding that a small number of users dominate the communication network. This research contributes to our understanding of how new social media technologies influence political communication, particularly among younger audiences.


\section*{Introduction}
As new technologies in the social media environment are rapidly reshaping patterns of political communication \cite{asbury2021effect}, how people consume political information via social media changes dramatically. Twitch, the entertainment-oriented live streaming platform, provides an environment where we can observe the role of new technologies like ``live-streaming'' and ``streaming-chat'' in shaping patterns of political communication. Contrary to popular belief, various political actors ranging from politicians to political pundits use the platform politically. Alexandria Ocasio-Cortez, an American politician serving as the U.S. Representative for New York's 14th congressional district since 2019, has live-streamed on Twitch several times since 2020. 

The popularity of Twitch streamer Hasanabi's live streaming of the 2020 US presidential election also garnered attention from journalists \cite{lor2020hasan}. The success of individual broadcasters like Hasanabi compared to traditional news media outlets like Fox News raises questions about why people choose to consume news content provided by individuals on platforms like Twitch. 

Besides, the unique technological features of Twitch, including real-time interaction, video and audio focus, and streaming chat, establish a distinct relationship between content creators and audiences, necessitating political communication research on the platform. This difference could affect the patterns of political communication on the platform, as content creators on Twitch would have more credibility and power due to their ability to interact with their audience in real-time \cite{mar2015you,mun2020you,sen2013mic}. The active usage of the streaming chat function grants more credibility by enabling reciprocal interaction among content providers and audiences. As they could communicate with the one who produces political messages, the audience would regard the political streamers as more relatable and accountable compared to political figures on other social media platforms \cite{lewis2020news}. At the same time, Twitch streamers rely heavily on their audience for financial revenue and content creation, more so than other social media platforms. The lack of active reactions from viewers during a political stream on Twitch can make it less appealing \cite{sjo2017twi}. Overall, the platform's real-time interactive nature, video and audio orientation, and streaming chat shape the environment of political communication by conditioning the credibility of political messages and granting audiences considerable agency over content production. By studying Twitch, we can understand better how these technological changes in social media can affect political communication in new social media.

The rise in political activities on the platform also calls for more systematic research on Twitch politics. Despite Twitch being an entertainment-oriented platform, political activities are increasing, with streamers speaking up on political issues such as Black Lives Matter \cite{lop2020blm,oca2020blm} and Capitol riots \cite{per2021cap}. Streamers who devote most of their broadcasting time to political issues are also present on Twitch. Another interesting feature is the composition of the users. As the most of platforms' content is mainly video-game related, most of the users in the platform tend to be younger compared to other major social media platforms, such as Twitter and Facebook. While more than 70\% of users fall into the 16-34 age group, there is still some portion of the older population \cite{dean_2021}. While 17\% of users are in the 35-44 age group, 10\% of users are in the age group above 45. As generational dynamics have been salient in recent American politics and other countries, such as South Korea or Japan, investigation of political communication inside the social media platforms where the composition of users is largely skewed toward younger generations can expand our understanding of generational politics and political communication in general.

To study the patterns of political communication in Twitch which has been understudied despite its theoretical and practical significance, I aim to answer three questions on Twitch politics in this paper by using various computational methods.

\begin{enumerate}
    \item \textit{Who are political Twitch streamers?} As Twitch is an extremely entertainment-oriented platform, the identification of political actors is essential to study Twitch politics. I found 574 political streamers using Twitch API and supervised machine learning techniques.
    \item\textit{What contents are covered in the political streams?} After the identification of political streamers, what kinds of political content are covered in the streams should be studied as we do not know what kind of political content is present in the platform and how political streamers cover it. 
    By collecting chat posts of political streamers and fitting topic models, I found political topics covered in the platform can be categorized into seven broad categories and a Twitch-specific pattern of political communication --- the usage of context-specific ``emotes (Twitch version of emoji)". 
    \item\textit{How do the political streamers and their audiences interact with each other?} As a live streaming platform with a very easy-to-use streaming chat function, Twitch facilitates interaction among streamers and audiences. The patterns of political communication among Twitch users, including both streamers and audiences, cannot be properly understood without properly capturing communication networks. I create reference networks of audiences of political streams to observe how they interact with one another and found most of the networks have a structure of opinion leadership.
\end{enumerate}

In the next section, I would introduce Twitch since the platform itself can be unfamiliar to some readers. Then, I describe why should we study Twitch politics. Next, I introduce why political actors choose Twitch as their means to deliver political messages with an emphasis on streaming chat. After that, I introduce three questions to understand Twitch politics and suggest the answers to those questions step by step.

\section{What is Twitch?}
Twitch is about a ten-year-old platform with 140 million unique worldwide monthly visitors and 41.5 million unique US monthly visitors in 2021 \cite{dean_2021}. The popularity of the platform looks more impressive when we compare it with traditional media: its monthly viewership in 2018 are comparable to those of some traditional cable TV networks in the US \cite{gil2018twi}. 
\begin{figure}[h!]
    \centering
    \includegraphics[scale=0.3]{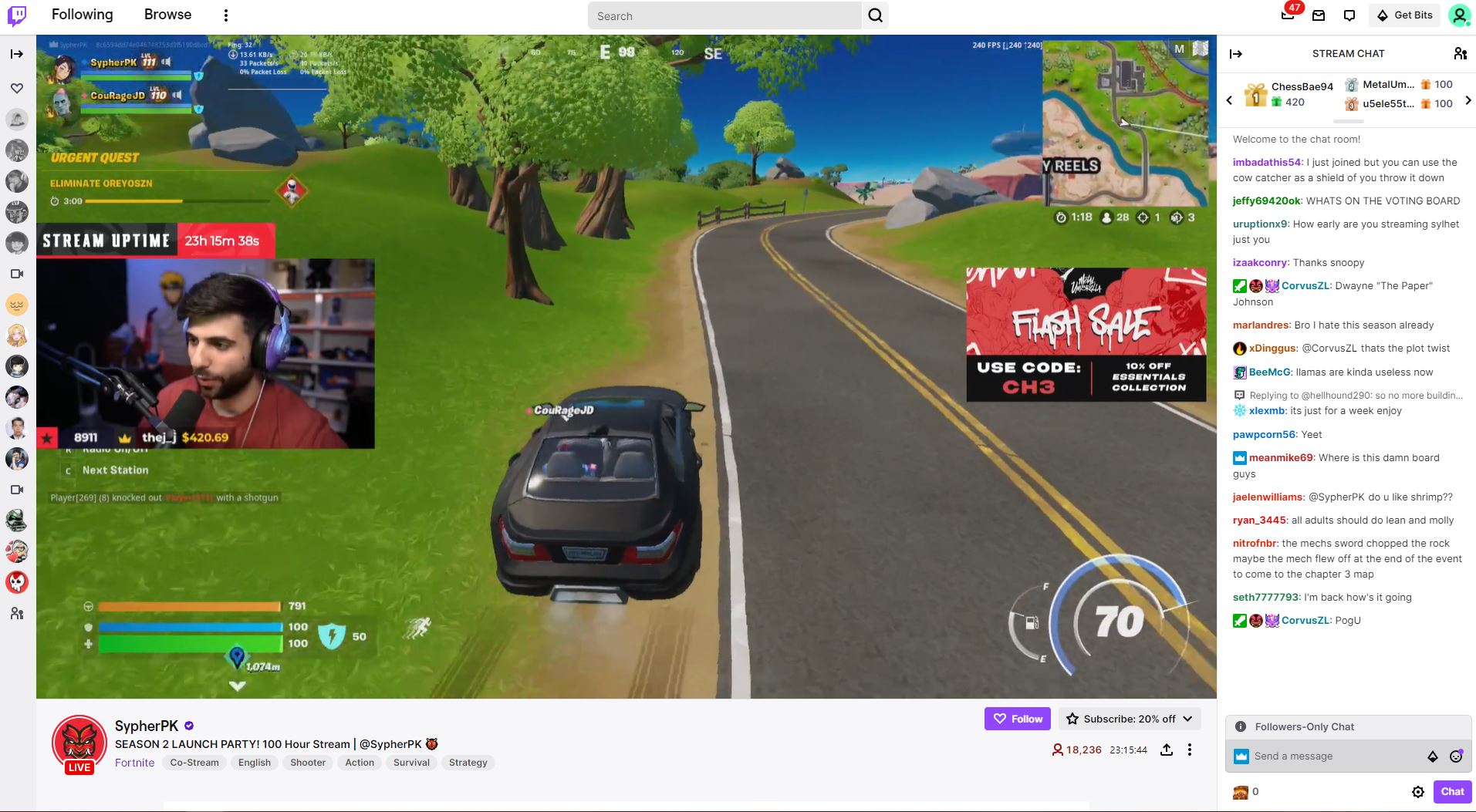}
    \includegraphics[scale=0.3]{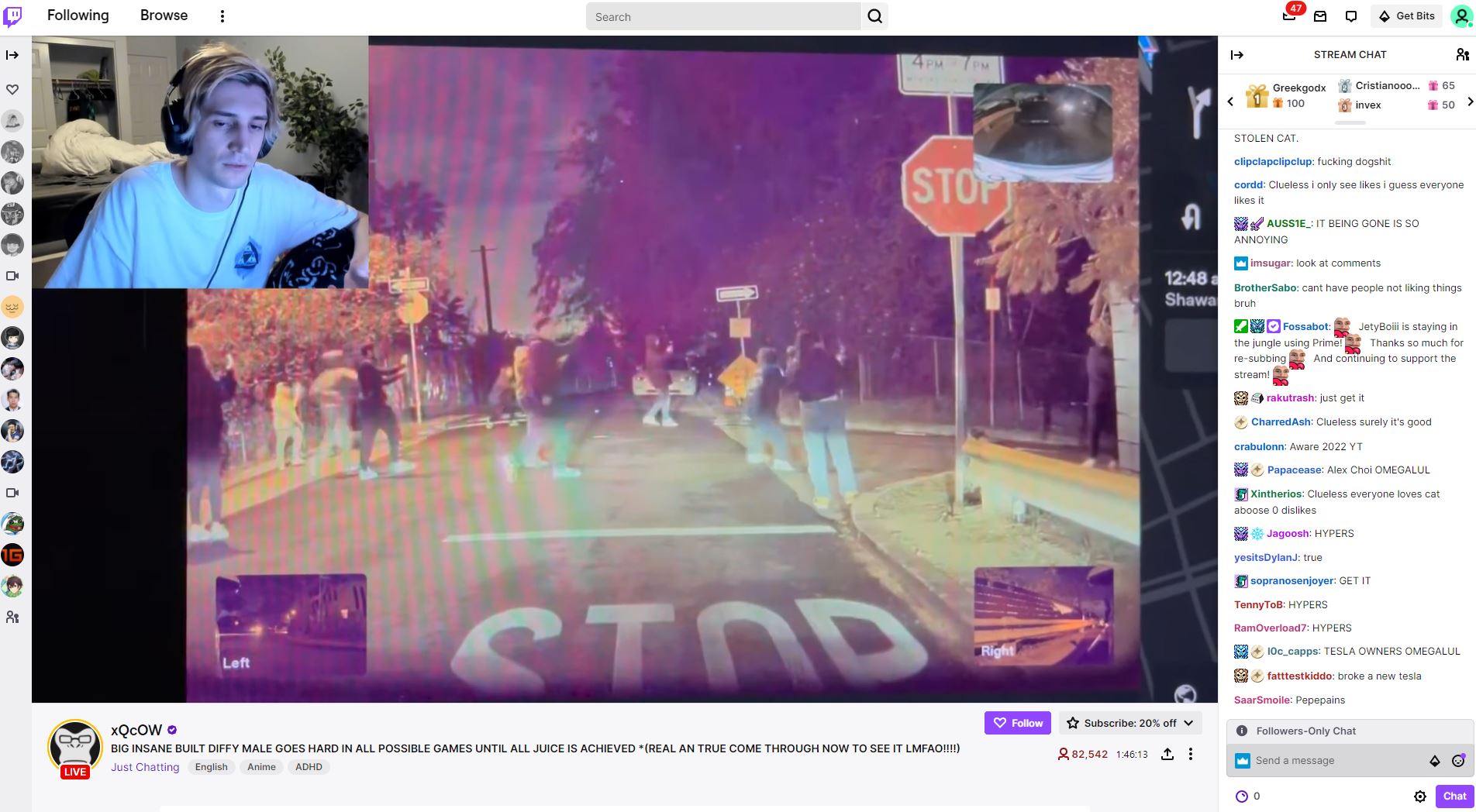}
    \caption{Screenshots of Twitch stream (Captured on 22/03/20)}
    \label{fig:twi_sch}
    \end{figure}
Figure~\ref{fig:twi_sch} shows the screenshots of Twitch streams. The upper image shows a screenshot of the typical gaming streaming of a Twitch streamer. He uses a microphone and webcam to show their appearance and voice while streaming their gaming screen, with a chat on the side for viewers to post messages. They sometimes watch videos or read texts with their audience and discuss the material together, as the lower image shows.

Several scholars from multiple disciplines have investigated Twitch from different angles ranging from relatively more macro-level approaches \cite{ask2019politics,johnson2019imperative} to studies that focus on technology-oriented traits \cite{churchill2016modem,claypool2015measurement} or Twitch affordances \cite{flores2019audience,jackson2020understanding,sjoblom2019ingredients}, such as streaming chats \cite{ford2017chat,hamilton2014streaming} and monetization \cite{johnson2019and}. Macro-level studies on Twitch as a platform economy point out that the flourishing of the platform has been possible because of both demand and supply for streaming \cite{ask2019politics,johnson2019imperative}.

Other scholars examine the platform more in detail. Some focus on the Twitch technologies themselves \cite{claypool2015measurement} and its influence on user network \cite{churchill2016modem}, while others address the importance of social affordances of Twitch \cite{jackson2020understanding,sjoblom2019ingredients} arguing the social structure surrounding those technologies also matters. Sj{\"o}blom et al. (2019) thoroughly examines Twitch's affordances while stressing the important role of webcams and audio as they are playing a huge role to establish bonds between streamers and their audiences. Similarly, Jackson (2020) also addresses the power of the synchronous nature that strengthens the perception of intimacy toward streamers, which is the root of their digital persona. 
Out of various Twitch affordances, the streaming chat function has received the most scholarly attention \cite{flores2019audience,ford2017chat,hamilton2014streaming}. Streaming chat is essential for audience engagement in live streaming and allows viewers to communicate with the streamer and other participants in the same stream, creating a community that can act as a virtual third place for Twitch users. 
Affordances related to monetization have been seriously considered by scholars as well \cite{johnson2019and,sjoblom2019ingredients} as they are the basic building block that buttresses the Twitch platform economy by providing monetary incentives to pursue streaming as a professional career \cite{ask2019politics,johnson2019imperative}. 

The political dynamics on Twitch have been studied with a focus on gender politics \cite{ruberg2019feeling, ruberg2019nothing, ruberg2021obscene, ruberg2021livestreaming, taylor2018watch} and the role of ``emotes" in live streaming political events \cite{riddick2022affective}. Research has highlighted the presence of misogynistic elements within the video game live streaming environment, influenced by policies on sexual content \cite{ruberg2021obscene} and community guidelines regarding the appearance of female streamers \cite{ruberg2019feeling, taylor2018watch}. These factors contribute to the derogation of female gaming streamers \cite{ruberg2019nothing} and impose additional demands for emotional labor related to affection \cite{ruberg2019feeling, ruberg2021livestreaming}. Furthermore, the use of ``emotes" in the context of political events has been examined, revealing their use as a tool for distracting other viewers \cite{riddick2022affective}.

Building upon the existing literature, this study aims to comprehensively identify political actors on Twitch and analyze their behavior, particularly focusing on the dynamics within streaming chats. By doing so, it seeks to provide a deeper understanding of the political landscape on Twitch, highlighting its significance and the implications of these interactions, which will be further explored in the following section.

\section{Why Twitch?}
Why should social scientists care about Twitch, a platform primarily known for its entertainment orientation? While there has been some research on Twitch politics, its political dimensions deserve further exploration. In this section, I will introduce a few episodes that underscore the importance of studying Twitch politics. The first and most famous episode is a series of streams by Alexandria Ocasio-Cortez, also known as AOC. Alexandria Ocasio-Cortez's series of gaming streams on Twitch in October 2020, which aimed to get out the vote for the incoming presidential election, was a huge success, with peak viewership of over 400,000 viewers \cite{riv2020aoc}. A later gaming stream with Canadian MP Jagmeet Singh and other Twitch streamers raised \$200,000 for charity \cite{hol2020aoc, dzh2020aoc}. Meanwhile, in her latest stream on Twitch, Alexandria Ocasio-Cortez discussed a political issue - the GameStop stock and Robinhood app issue - with guests including Alexis Goldstein, Alexis Ohanian, and TheStockGuy. They expressed their opinions on the issue and the financial system in general, with AOC advocating for systematic financial reform in the US. The stream was successful, with around 200,000 peak viewers.

Other than politicians, there are other political actors in Twitch: Twitch streamers who stream political content. The best example of a renowned political streamer would be ``HasanAbi". 
His US 2020 presidential election marathon stream on November 3rd, 2020 for 16 hours had 225,000 viewers at its peak \cite{lor2020hasan}. Viewership of his election day Twitch stream was even comparable to the major media outlets when we use ``Total Hours Watched" as a criterion to compare what media outlets people have chosen on election day: the stream's portion (4.9\%) is almost similar to that of Fox News (6.5\%) \cite{cal2020hasan}. The huge success of the election day stream by HasanAbi invited the attention of various news media outlets on the potential of Twitch as a political conduit, along with the huge success of AOC's few streams. As he clearly states in the interview with New York Times, his intention of the Twitch stream has been mainly political - he wanted some platform to congregate people every day and deliver his political opinion \cite{lor2020hasan}.
Even entertainment-oriented streamers sometimes speak up about political issues, with a notable example being the Black Lives Matter protests in mid-2020. Some streamers expressed their opinions during live streams or produced videos related to the issue, while others held fundraising for Black Lives Matter \cite{lop2020blm}. Twitch politics is not confined to political streamers or viewers, but a broader pool of users and streamers on the platform.

Theoretically, studying Twitch politics grants a novel opportunity to understand how new technologies in the social media environment affect the patterns of political communication. Twitch, an entertainment-focused live streaming platform, offers an environment for us to study how emerging technologies such as ``live-streaming" and ``streaming-chat" influence the dynamics of political communication.

\section{What makes political content creators choose it?: The importance of streaming chat function and systematic dependence on audiences}
People who want to create political content on a streaming platform may choose Twitch because it provides an easy environment to start streaming. Creating video content for YouTube can be efficient compared to producing text content \cite{mun2020you}, and live streaming on Twitch is even more efficient because the time spent for streaming is the same as the length of the show. The user-friendly interface of Twitch and the availability of the app ``Twitch Studio" make it easy for anyone to start streaming using their PC or mobile phone. The only requirements are an electronic device to run the program and objects to show viewers, primarily the streamer themselves.

As the previous literature thoroughly examines, the existence of various ways to pursue financial revenue also makes Twitch more attractive \cite{johnson2019and,sjoblom2019ingredients}. Twitch offers direct revenue streams such as subscriptions, viewer donations through Twitch currency called Bits, and advertisement revenue. Streamers can also earn money through external sources such as sponsored product placement and third-party donation platforms. Although not all streamers can earn a stable income, the potential for financial rewards is a factor that attracts users to the platform.
Hence, Twitch provides an ideal platform for political content creators to brand themselves as micro-celebrities. By using webcams and microphones \cite{sjoblom2019ingredients} to create a personal connection with their audience, they can establish credibility \cite{mar2015you} and foster a sense of community through parasocial relationships \cite{sherrick2023parasocial}. This connection can be further strengthened by including links to other social media platforms in their profiles \cite{sjoblom2019ingredients}, enhancing their overall engagement with the audience.
The chat function on Twitch enables the communication between streamers and viewers, creating a space for mass communication that was not possible in traditional media outlets \cite{mun2021str}. Viewers can instantly react to what the streamer says or does by using Twitch emotes or normal languages, which helps create a relatable and accountable relationship between the streamer and the audience \cite{lewis2020news}.

Twitch's real-time chat function provides an attractive feedback loop for political content creators and audiences to have a discussion on political topics in real-time, which can be very attractive to political content creators. While platforms like Twitter and Facebook allow for discussion on political topics, there is usually a time lag between users' responses. However, Twitch's chat function allows political content creators to express their opinions in real-time while engaging with their audience, giving the impression of a live discussion. This engagement can attract potential viewers interested in expressing their opinions on contentious political topics. Overall, the effective streaming chat function which enables political streamers to have a real-time discussion with their audiences can be the most important feature of Twitch politics. Exploring the role of streaming chat on political communication in the social media context can have some implications for the political communication literature as we little know how it affects inter-user communication patterns.

The other important trait of Twitch is that the creation of the content is also dependent on viewers at the same time. Not to mention the audience is a source of revenue that streamers are making \cite{johnson2019and}, their content creation itself is also inherently dependent on viewers and their chat posts because of the real-time nature of live streaming \cite{sjo2017twi}. Viewers' reactions during live streaming are crucial for political content creators, as their content tends to be talk show-oriented and requires smooth communication with viewers through streaming chat. Without active participation from viewers, political streaming cannot be successful and will only be a manifestation of the creator's political opinion. The audience users of Twitch have powerful agency and their participation is essential for the success of streamers. This highlights the need to study the audiences of political streams and their interactions with each other, not just the political streamers.


\section{Three Questions: Who are they, What do they broadcast and How do they interact?}
In this paper, I answer three questions on Twitch politics. First, who are the political streamers on the platform? To study Twitch politics systematically, it is imperative to identify political streamers first from the population of streamers, who are mostly non-political. I have retrieved extensive lists of streamers who may stream political content and their information using Twitch API. Based on the information, I have leveraged supervised machine-learning techniques to find political streamers from the retrieved lists.

Second, what do they broadcast during their streaming? It is important to conduct a study on the types of political content that are discussed in streams, since we are unaware of the political content that exists on the platform and how it is addressed by political streamers. Are the topics that are widely covered by traditional media also get covered by streamers? Or, do they shed a light on other topics that are not fully covered by other media outlets? Observation of their topic choices can help us to comprehend the basic patterns of Twitch politics. 
I will answer the question by conducting various text analyses on the originally collected streaming chat post data.

Third, how do political actors, both political streamers and their audiences, interact with one another? In order to fully comprehend the political communication behaviors of Twitch users, which includes both streamers and their audiences, it is necessary to accurately capture the communication networks. 
I would focus on inter-audience communication in a stream that is done by mentioning others' usernames. 
How do Twitch users in political chat rooms behave in terms of communicating with each other via streaming chat? By constructing and analyzing reference networks of each political stream, I will study political communication networks in Twitch political streams.  

\section{Question 1: Who are the political streamers?}
In this section, I find political streamers out of extensive lists of streamers that are retrieved via Twitch API using the supervised machine learning method. In the first subsection, I briefly introduce Twitch API and how I have utilized it to get extensive lists of streamers. Then, I introduce the coding scheme to classify political streamers and the classified list of political streamers.

\subsection{Using Twitch API to retrieve extensive streamer lists}
I have collected a week of streaming records from 21-08-31 00:25 EST to 21-09-08 16:44 EST of Twitch broadcasting with the three game-names, ``Just Chatting", ``Talk Shows and Podcasts", and ``Politics", which are identified through the process fully described in S1, using Twitch API. 
The streaming records contain various information including user name, user id, streaming title, etc. I identified 53,550 unique ids for ``Just Chatting", 5,853 for ``Talk Shows and Podcasts" and 324 for ``Politics". As I have also collected Twitch profiles 
of those streamers as well, I merged broadcasting titles of streamings for a week and text information in their streaming profiles to create text data of all streamers. 

\subsection{Finding political streamers using supervised machine learning techniques}

I have used a supervised machine-learning technique to identify political streamers from the user data I have collected. The political streamers are defined by the following coding rules refer:
\begin{enumerate}
    \item If a streamer has broadcasted political content at least once (i.e. ``Texas Abortion Law is a shame"), I coded her as a political streamer. 
    \item If a streamer profile shows that she broadcasts about or is at least interested in politics (i.e. ``I sometimes talk about politics"), I coded her as a political streamer. \item If a streamer explicitly identifies their political interest or partisanship on her profile (i.e. ``I am Leftist", ``This is conservative podcast"), I coded her as a political streamer. 
    \item If a broadcasting title or streamer profile contains a representative term or hashtags of specific political movements (i.e. ``\#BLM", ``\#FreePalestine"), I coded her as a political streamer.
\end{enumerate}\par
\begin{figure}[h!]
    \centering
    \includegraphics[width=0.8\linewidth]{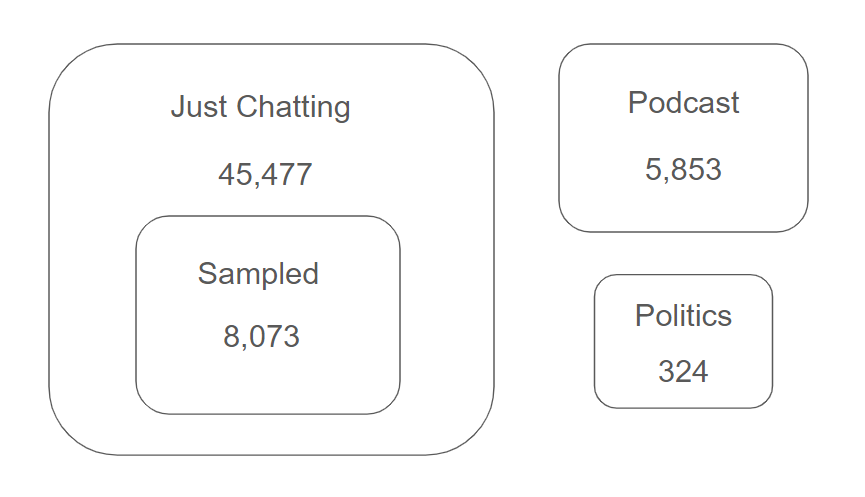}
    \caption{Venn Diagram of Twitch Streamer Data}
    \label{fig:venn}
\end{figure}

As I already know that there would not be as many political streamers in the list of ``Just Chatting" compared to ``Talk Shows and Podcasts" and ``Politics" as described in S1, I adopt the following strategy to produce training sets with enough target values, political streamers, to ensure the model gets trained properly. First, I have hand-coded all components from the ``Talk Shows and Podcasts" and ``Politics" sets in Figure~\ref{fig:venn}. Then, I extracted the top 30 words from the combined text data of political streamers found in the first step and filtered out nonessential words, such as articles and be-verbs, and other common words that are widely believed to be frequently used by most Twitch streamers (i.e. stream, live, etc.). I have used these keywords as a filter to sort out observations that are more likely to be coded as political from ``Just Chatting" data. The process gives the ``Sampled" set with 8,037 streamers that is right inside of the ``Just Chatting" set in Figure~\ref{fig:venn}. I hand-coded all of these streamers and identified 225 political accounts. The whole process gives me a total of 14,250 hand-coded train data with 550 political streamers. By using the labeled data, I fitted the supervised machine learning classifier and was able to identify 574 political streamers in total. Details about the machine learning model training process are described in S2.\

\section{Question 2: What do political streamers broadcast?}

\subsection{Collecting Chat Posts from political streamers' live streaming using 'chat bots' - Descriptive Figures}
As a first step to studying political discussion of identified political streamers, I have used chatbots that get connected to each stream and exist in the channel until it gets offline \cite{flores2019audience} to collect chat posts from streams provided by political streamers I have identified. The bots allowed me to download all chat posts in the stream's associated IRC (Internet Relay Chat) with various information, ranging from the text of chat posts to the user-name of the sender \cite{flores2019audience}. Using the pipeline, I have collected chat posts of political streamers from 2021-12-11 
to 2022-03-25.

Due to the time lag between the identification process of political streamers and actual chat post data collection, there has been some loss in the number of political streamers. I was able to collect chat posts of 478 political streamers out of a total of 574, which means the survival rate is above 83\%. The reasons for the loss can be diverse. Some streamers might have quit streaming at all, or some might just want to cease it for a while.

\begin{figure}[h!]
    \centering
    \includegraphics[width=0.5\linewidth]{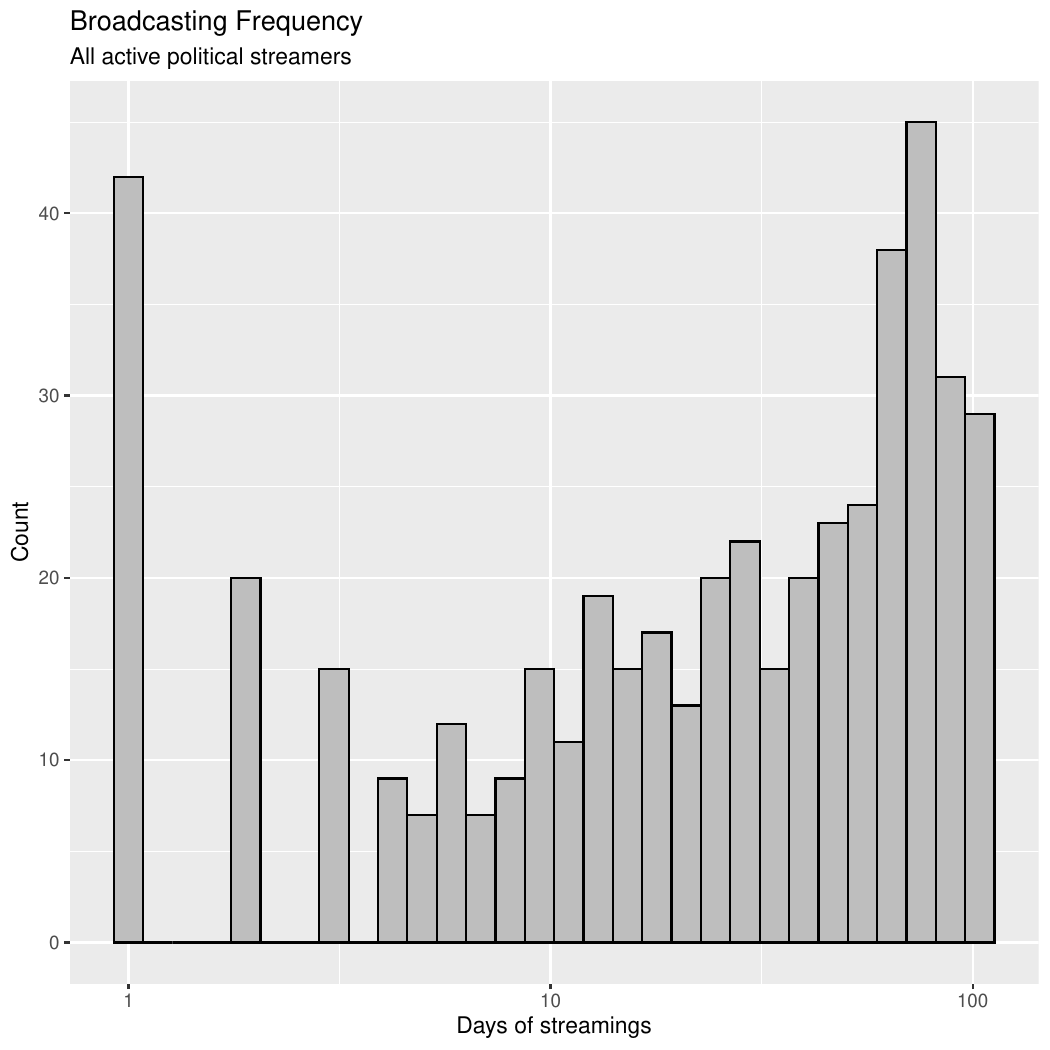}
    \caption{Frequency of streamings by political streamers}
    \label{fig:freq_stream}
\end{figure}
Figure~\ref{fig:freq_stream} illustrates the daily frequency of political streamers starting their streams during the data collection period, which spanned from December 11, 2021, to March 25, 2022. Since this period covers 105 days, the maximum value in the plot is naturally 105. The distribution is right-skewed, indicating that while some streamers were very active, others were less so. Specifically, 342 political streamers streamed on more than 10 days, and 228 of them streamed on more than 30 days over the three-month period. This suggests that most of the data captures regular interactions between viewers and streamers, although some streamers broadcasted infrequently.

\begin{figure}[h!]
    \centering
    \includegraphics[width=0.45\linewidth]{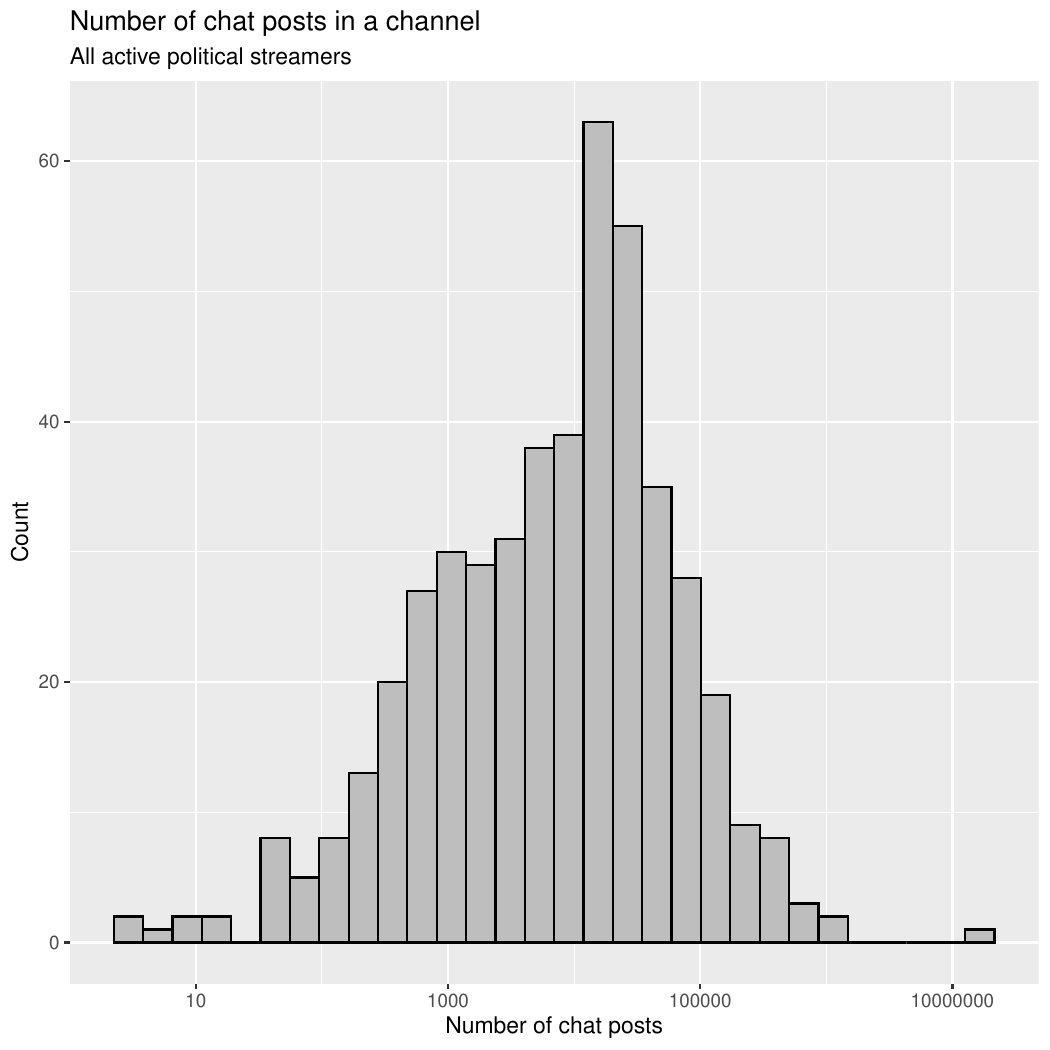}
    \includegraphics[width=0.45\linewidth]{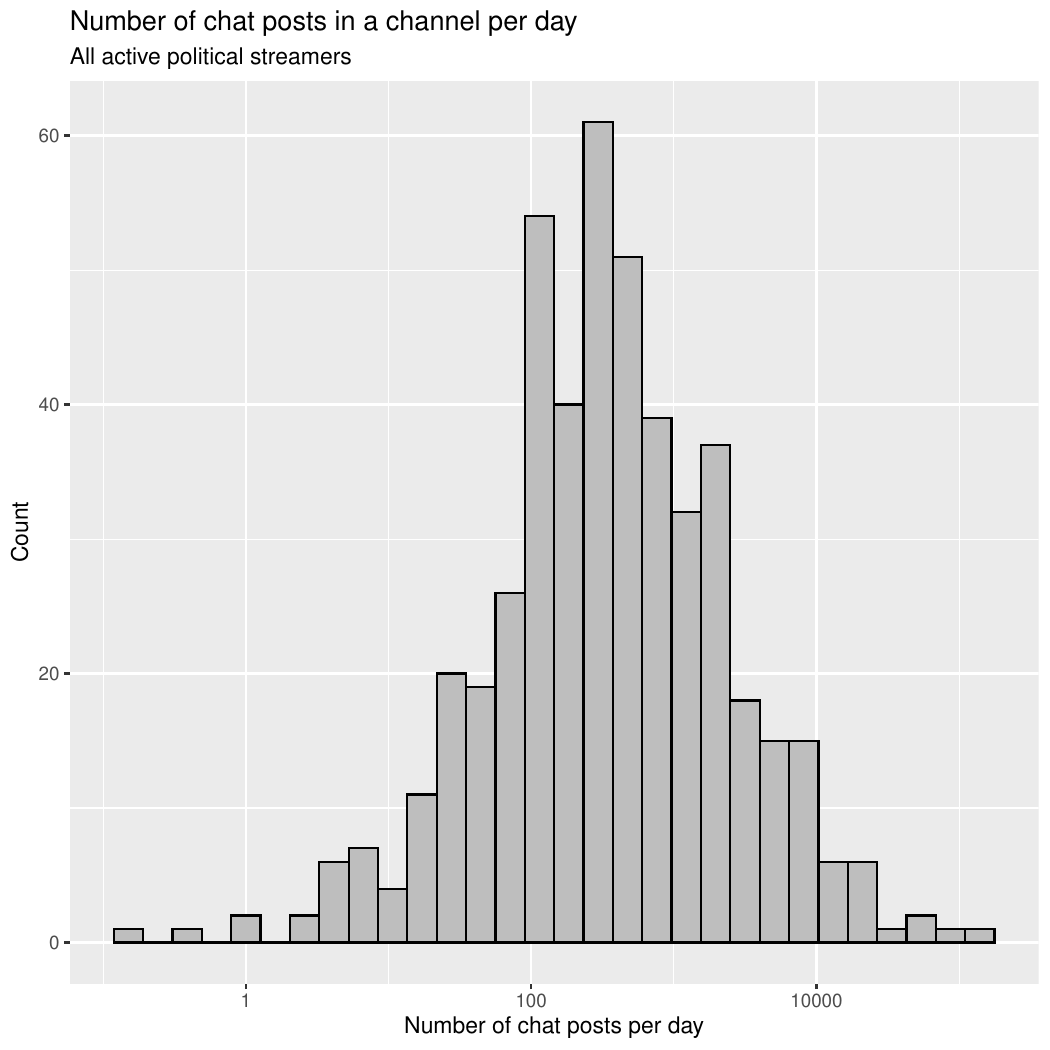}
    \caption{Number of chat posts. The x-axis is log transformed while labels show the raw values.}
    \label{fig:freq_chat}
\end{figure}
Figure~\ref{fig:freq_chat} shows the number of chat posts on each streamer's channel, both in total and on a daily basis. The logged scales of both histograms approximate a normal distribution, with a small number of extreme values on both ends. The total number of chat posts collected from all channels is 33,649,628. The channel with the most chat posts has 16,945,559, while another channel has only 2 posts from a single viewer. The mean and median values are 70,396 and 2,430, respectively. On a daily basis, where the number of chat posts per channel is divided by the number of streaming days, the mean and median values are 2,224 and 339, respectively. This suggests that, for a political channel in the middle of the distribution, we can expect at least 350 chat posts per stream. This indicates that most political channels receive some level of audience interaction, reflecting feedback loops in the political streams.

\begin{figure}[h!]
    \centering
    \includegraphics[width=0.45\linewidth]{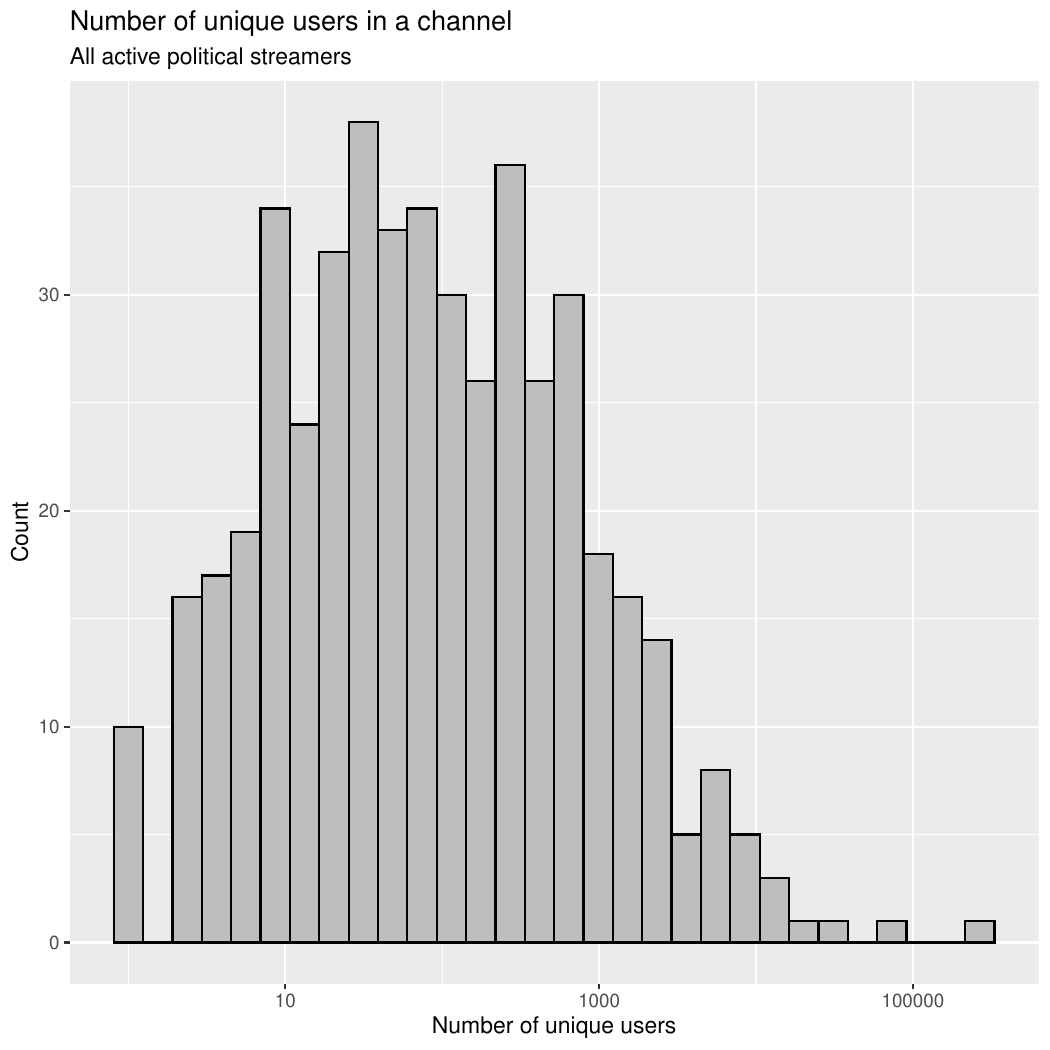}
    \caption{Number of unique users in a channel.The x-axis is log transformed while labels show the raw values.}
    \label{fig:freq_user}
\end{figure}
Figure~\ref{fig:freq_user} illustrates the distribution of unique usernames that have posted at least one chat post on a channel. The histogram, plotted on a logged scale, reveals a slightly right-skewed pattern. The most popular channel boasts 267,206 unique users, while the least popular has only one unique user. In total, there are 646,073 unique users across all channels, with mean and median values of 1,351 and 70, respectively. The higher mean compared to the median is largely influenced by extreme values on the right-hand side of the distribution. Therefore, most of channels are likely to have around 70 unique users who have posted at least one chat post during its streaming. While a smaller audience size can foster active political discussions among users and streamers, establishing parasocial relationships \cite{sherrick2023parasocial} and gaining micro-celebrity status \cite{mar2015you} may be challenging due to their dependence on popularity.

\begin{figure}[h!]
    \centering
    \includegraphics[width=0.45\linewidth]{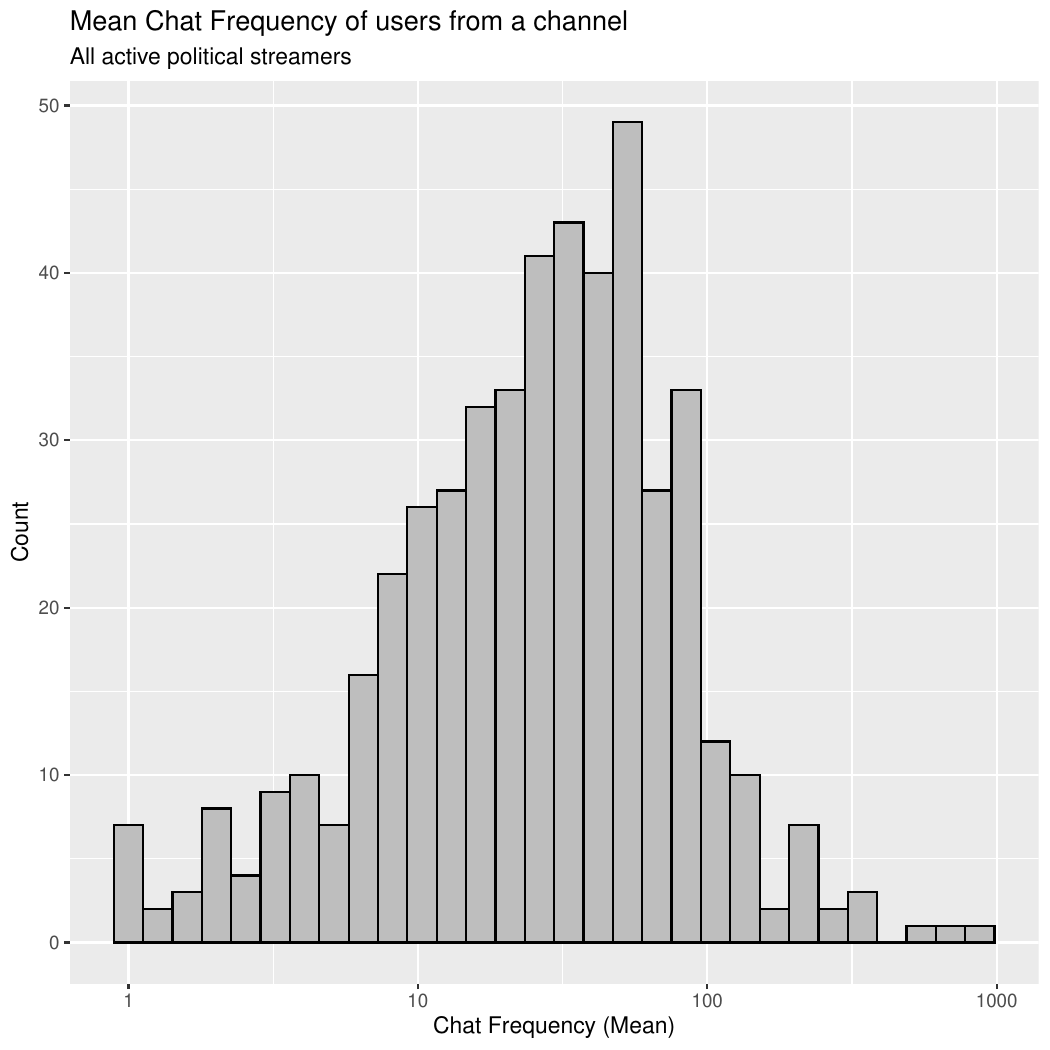}
    \includegraphics[width=0.45\linewidth]{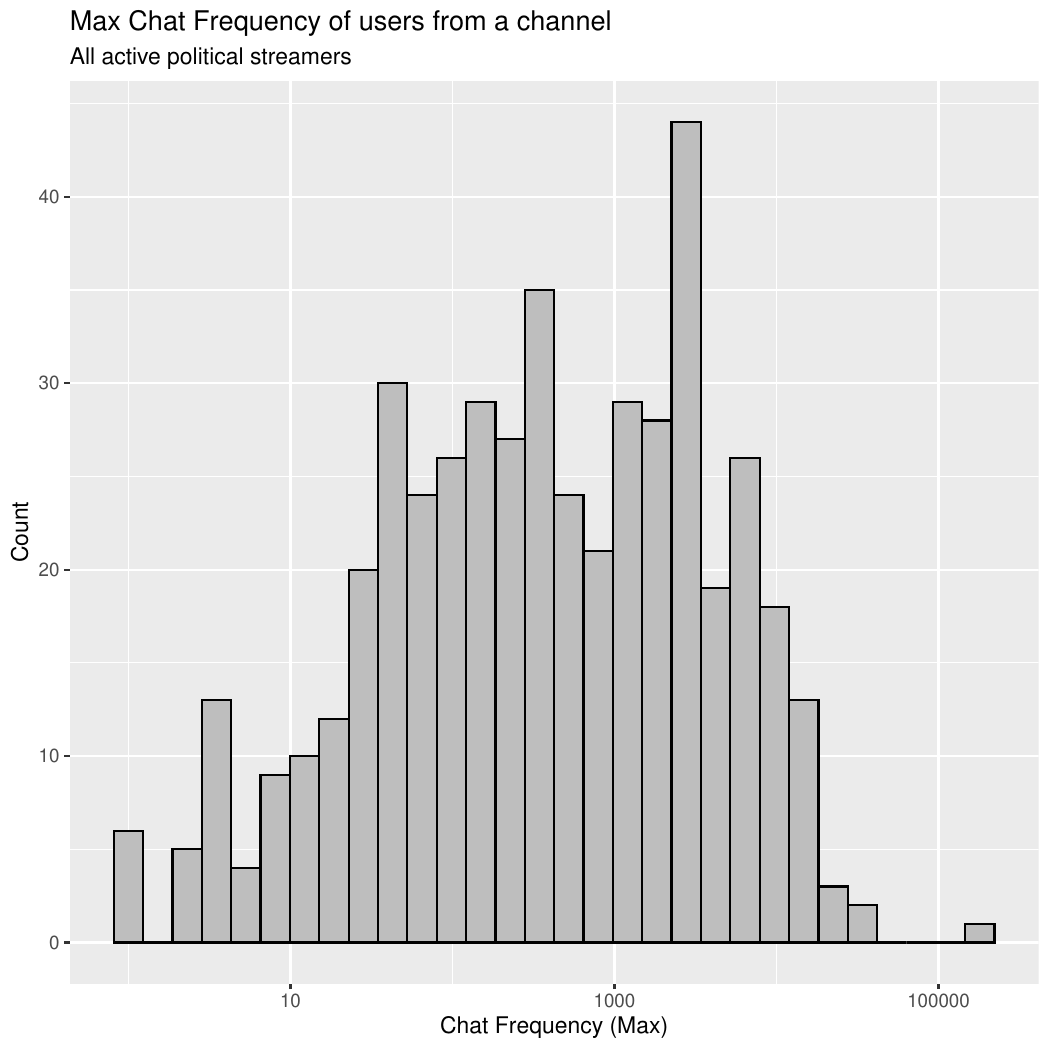}
    \caption{Frequency of chat posts}
     \label{fig:freq_chatpost}
\end{figure}
Figure~\ref{fig:freq_chatpost} shows the frequency of chat posts at a user level. The mean chat frequency of all users in a channel displays a nearly normal distribution, whereas the maximum chat frequency has a weak left-skew. The mean and median values for the mean chat frequency are 45.3 and 28.5, respectively, while those for the maximum chat frequency are 2,475 and 377. These patterns suggest that most users are participating in the chat repeatedly. However, the significantly high number of chat posts from a single user suggests the presence of bots, which are commonly used in Twitch streams.

\subsection{What is discussed in political Twitch streams?: Topic models of the chat post data}

In this section, I will discuss the content of political streams by conducting topic modeling on chat post data. Topic modeling, a tool widely used in both social science and computer science \cite{roberts2019stm, liebman2020mass}, helps us observe general trends in the text corpus by examining broad categories among chat posts. Through topic modeling, we can identify topics, which are sets of words that frequently occur together within documents \cite{liebman2020mass}.

Due to the high computational cost, I sampled 5\% of chat posts from each political stream. The number of chat posts collected varies significantly by stream because of differences in audience size and broadcasting frequency. Therefore, I sampled 5\% of chat posts from each channel instead of sampling at an aggregate level. The total number of sampled chat posts is 1,682,739. I identified a topic as political if one or more politically relevant keywords were among the top 15 keywords within a topic. Based on this criterion, I found 45 political topics out of a total of 150 topics. This means that 30\% of the topics identified through the analysis are political. A comprehensive list of political topics is provided in S3.

\begin{table}[hbtp!]
\caption{Topic Proportions. Proportion for political chat posts and all chat posts are calculated by dividing the number of topics with the number of political topics, 45, and the total number of topics, 150.}\centering\small
\label{tab:topic_proportions}
\begin{tabular}{|l|l|l|l|}
\hline\hline
Category  & Number of topics & Proportion (Political) & Proportion (Total) \\ \hline\hline
International issues  & 11 & 25\% & 7\%\\ \hline
US politics  & 10  & 21\% & 6\%\\ \hline
Identity politics  & 9  & 21\% & 6\%\\ \hline
Ideological debate  & 7  & 15\% & 4\%\\ \hline
Politics in general  & 4  & 9\% & 3\%\\ \hline
Public health and politics  & 2 & 4\% & 1\%\\ \hline
Environmental issues  & 2  & 4\% & 1\%\\ \hline
\end{tabular}
\end{table}

I have categorized political topics into seven categories based on the political keywords in those topics: 1) International issues, 2) US politics, 3) Identity politics, 4) Ideological debate, 5) Politics in general, 6) Public health and politics, and 7) Environmental issues. Table~\ref{tab:topic_proportions} shows the number of topics in each category and the proportion each category represents in the corpus of political chat posts and all chat posts. 

\begin{figure}[h!]
    \centering
    \includegraphics[width=1.0\linewidth]{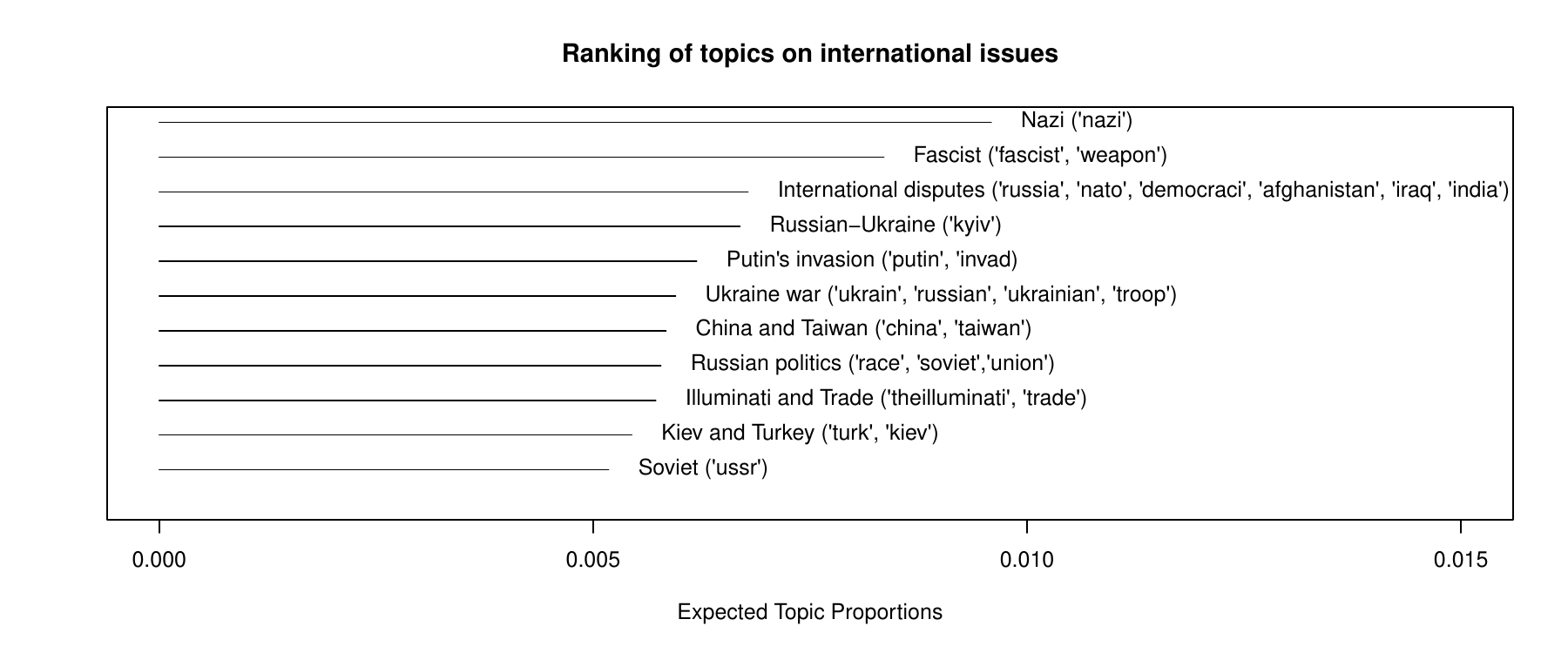}
    \caption{Topic Ranking: International issues}
    \label{fig:top_int}
\end{figure}

The most frequently appearing political topics in Twitch streams are international issues. These topics also occupy the largest proportion in both the political chat posts (25\%) and all chat posts (7\%). Figure~\ref{fig:top_int} provides more details about the topics on international issues. The prevalence of international issues on Twitch is largely due to the specific incident that occurred during the data collection period: the Ukraine invasion by Russia. Six topics are directly related to the war or Russia. Other topics address international disputes involving NATO-related countries (``International disputes'') and the China-Taiwan conflict (``China and Taiwan'').

\begin{figure}[h!]
    \centering
    \includegraphics[width=1.0\linewidth]{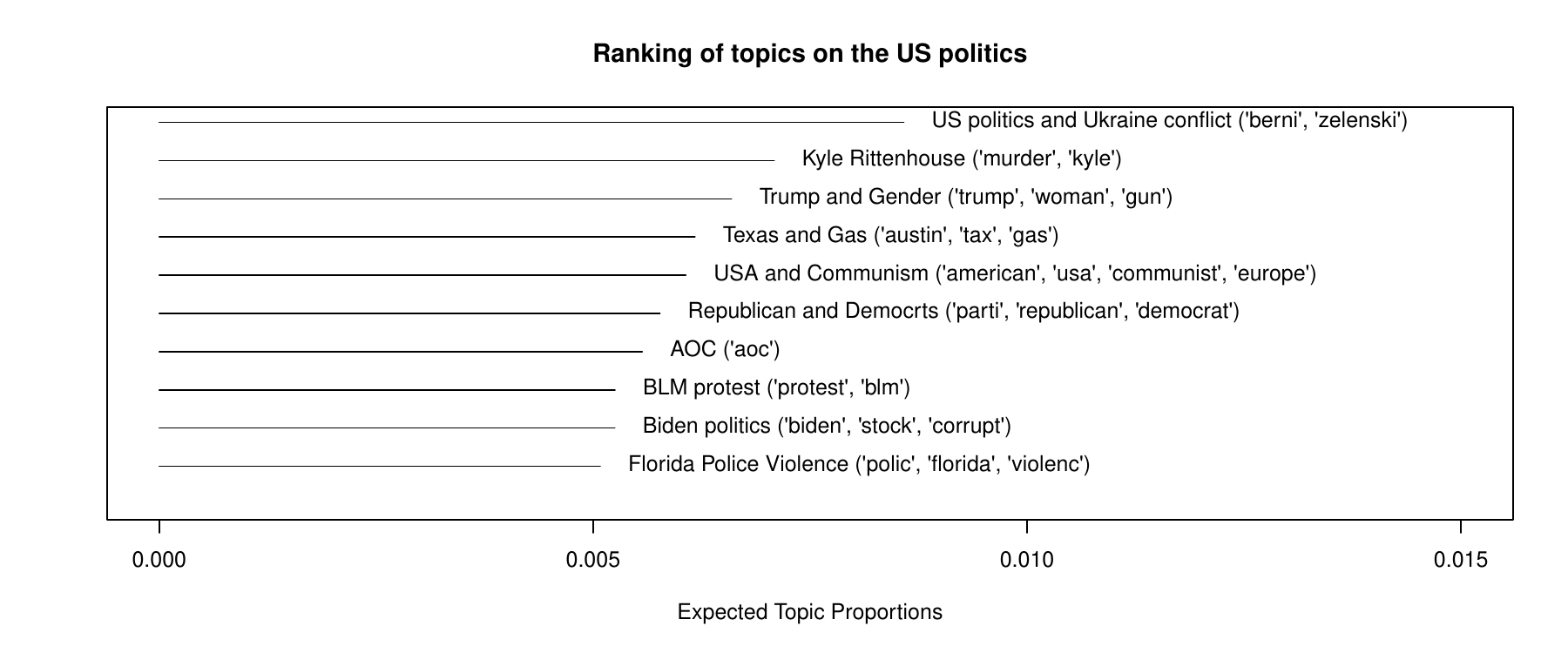}
    \caption{Topic Ranking: The US politics}
    \label{fig:top_us}
\end{figure}

As Figure~\ref{fig:top_us} shows, topics related to US politics also appear frequently: the number of topics is 10 and they account for 21\% of political chat posts and 6\% of all chat posts. As Twitch is the most popular in the United States, this finding is not that surprising. There is a topic about Trump (``Trump and Gender") and Biden (``Biden politics"), the two most popular politicians in the United States. Alexandria Ocasio Cortez was also mentioned in chat posts (``AOC"). Topics related to controversial violent events, such as Kyle Rittenhouse incident (``Kyle Rittenhouse") and police violence (``Florida Police Violence"), and the Black Lives Matter protest (``BLM") appear in political streams as well. 

As shown in Figure~\ref{fig:top_us}, topics related to US politics appear frequently. There are 10 topics, accounting for 21\% of political chat posts and 6\% of all chat posts. Given that Twitch is most popular in the United States, this finding is not surprising. Notable topics include discussions about Trump (``Trump and Gender'') and Biden (``Biden politics''), who are the two most prominent politicians in the country. Alexandria Ocasio-Cortez is also mentioned (``AOC''). Additionally, topics related to controversial violent events, such as the Kyle Rittenhouse incident (``Kyle Rittenhouse''), police violence (``Florida Police Violence''), and the Black Lives Matter protest (``BLM''), appear in political streams as well.

\begin{figure}[h!]
    \centering
    \includegraphics[width=1.0\linewidth]{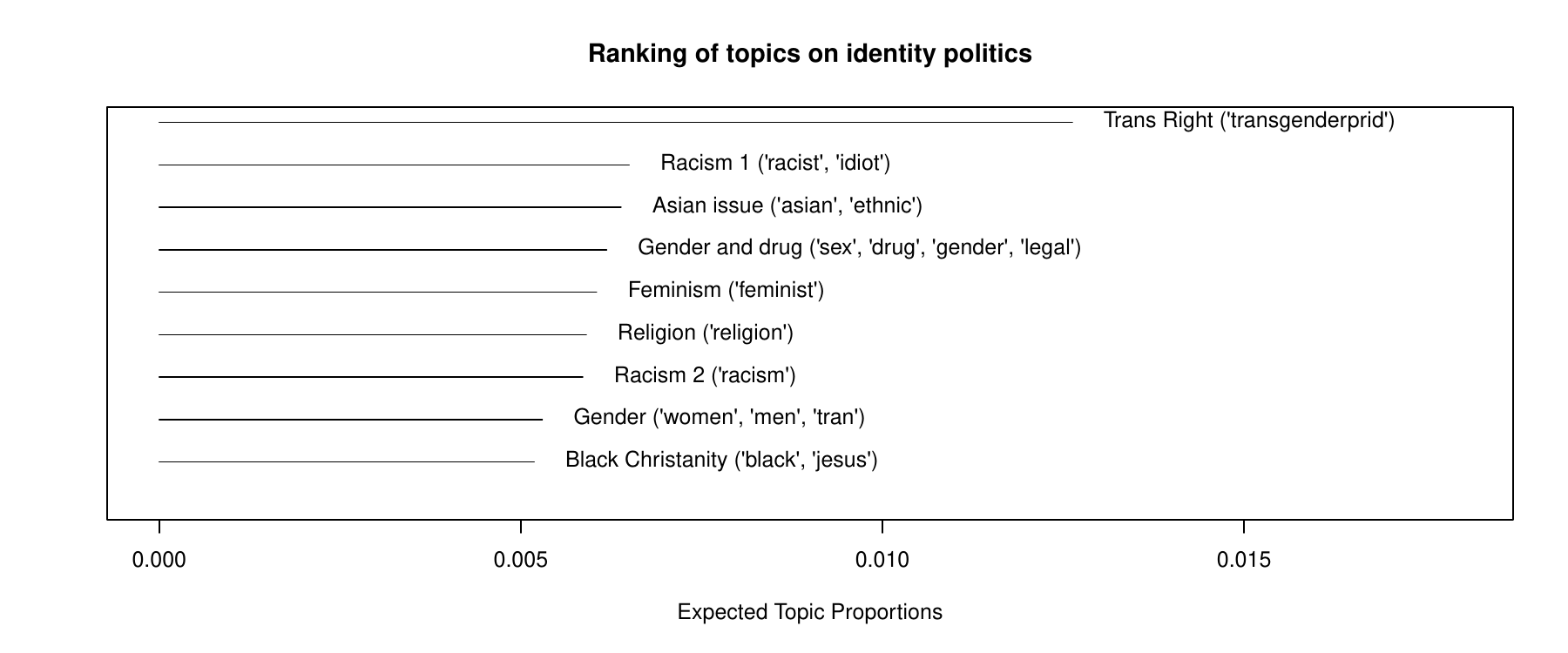}
    \caption{Topic Ranking: Identity politics}
    \label{fig:top_iden}
\end{figure}

Interestingly, topics related to identity politics appear almost as frequently as those related to US politics and international issues. There are 9 topics, accounting for 21\% of political chat posts and 6\% of all chat posts. Figure~\ref{fig:top_iden} illustrates the identity politics-related topics. The most frequently appearing topic in the political chat post corpus is related to trans rights (``Trans Right"). Gender issues are also widely covered, with three topics related to gender politics. Racism is another significant issue in political streams, with three topics addressing racism or race-related issues (``Asian issue"). This finding aligns with the literature, which suggests that Twitch and other online spaces with streaming chat functions can act as virtual third places, revealing users' identities \cite{hamilton2014streaming}, and fostering a sense of community through parasocial relationships between streamers and audiences \cite{sherrick2023parasocial}. Audiences may feel that Twitch political streams provide a space where they can reveal their identities and comfortably discuss related issues.

\begin{figure}[h!]
    \centering
    \includegraphics[width=1.0\linewidth]{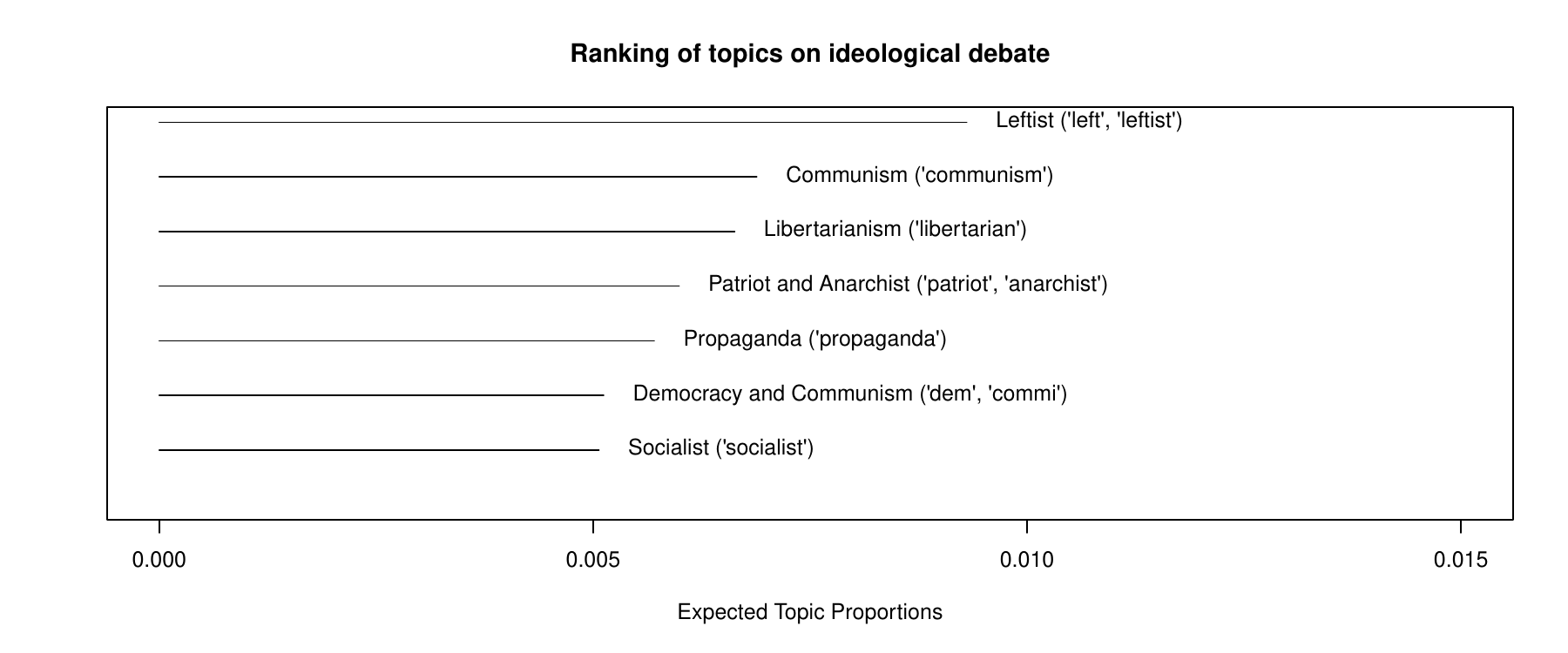}
    \caption{Topic Ranking: Ideological debate}
    \label{fig:top_ideo}
\end{figure}

Topics related to ideological standings are also present. There are 7 topics, comprising 15\% of political chat posts and 4\% of all chat posts. Figure~\ref{fig:top_ideo} provides more details about these topics of ideological debate. More than half of the topics reference leftist ideologies, while only two topics indirectly relate to conservative or right-wing ideas. This may suggest that right-wing-oriented political streams engage in self-reinforcing communication by condemning leftist political enemies. Additionally, the categories "Public health and politics" and "Environmental issues" each have two topics, accounting for 4\% of political chat posts and 1\% of all chat posts, respectively.

\begin{figure}[h!]
    \centering
    \includegraphics[width=1.0\linewidth]{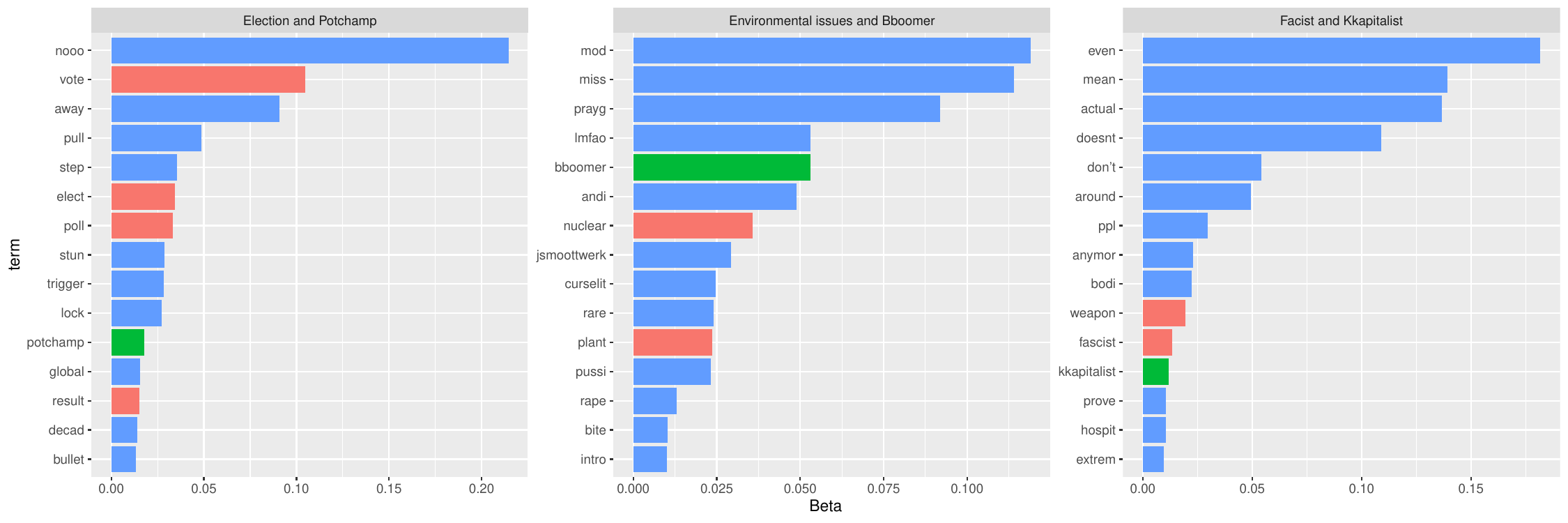}
    \caption{Emote usages in political streams}
    \label{fig:emo_freq}
\end{figure}

\begin{figure}[h!]
    \centering
    \includegraphics[width=0.2\linewidth]{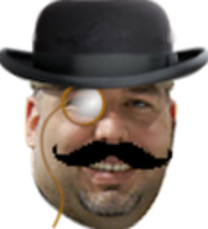}
    \includegraphics[width=0.2\linewidth]{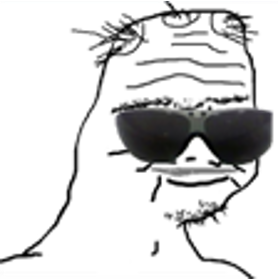}
    \caption{Kkapitalist and Bboomer}
    \label{fig:emo_img}
\end{figure}
One notable aspect is the use of Twitch-specific communication modes, particularly ``emotes," in political discussions. Twitch emotes are akin to emojis used in streaming chats. Figure~\ref{fig:emo_freq} demonstrates that three emotes—Potchamp, Bboomer, and Kkapitalist—are frequently employed in chat posts categorized as political. Figure~\ref{fig:emo_img} depicts two commonly used emotes, Kkapitalist and Bboomer, designed to satirize affluent capitalists and Baby Boomers, respectively. The prevalence of these emotes suggests certain political inclinations among Twitch users, potentially indicating anti-capitalist and anti-Boomer sentiments. Alternatively, this usage can be interpreted with more context. For instance, some users may associate Boomers with environmental concerns, as keywords related to environmental issues often co-occur with ``Bboomer" in the topic Environmental issue and Bboomer in Figure~\ref{fig:emo_freq}. Overall, political discourse on Twitch utilizes not only text but also context-specific emotes, which merits further exploration by scholars of political communication.

\section{Question 3: How do they interact?}
\subsection{Reference network in chat posts within a political stream}
Using the collected chat post data, I have constructed 'reference' networks of chat posts from each political streamer to observe user interactions. In Twitch streaming chats, users often refer to streamers or other users for various purposes, such as sending emotes, asking related questions, or engaging in casual conversation. This section explores how Twitch users in political streams communicate with each other through streaming chat and examines the types of content they share. Specifically, I illustrate the structure of user reference networks within political streams and highlight insights gained from analyzing these interactions.

I identified all unique users who posted at least one chat post in each channel and considered them as nodes for reference networks. The edges in these networks were established through the following steps: 1) Identifying chat posts containing `$@$' symbols. 2) Extracting the subsequent words after the symbol as references to other users. 3) Establishing directed edges from the user who posted the chat to the mentioned user if their name matched the extracted word. Out of 478 active political streamers, I created 279 directed reference networks for channels where at least one chat post mentioned another user within the same stream.

\subsubsection{Network Descriptive Statistics}

\begin{figure}[h!]
    \centering
    \includegraphics[width=0.4\linewidth]{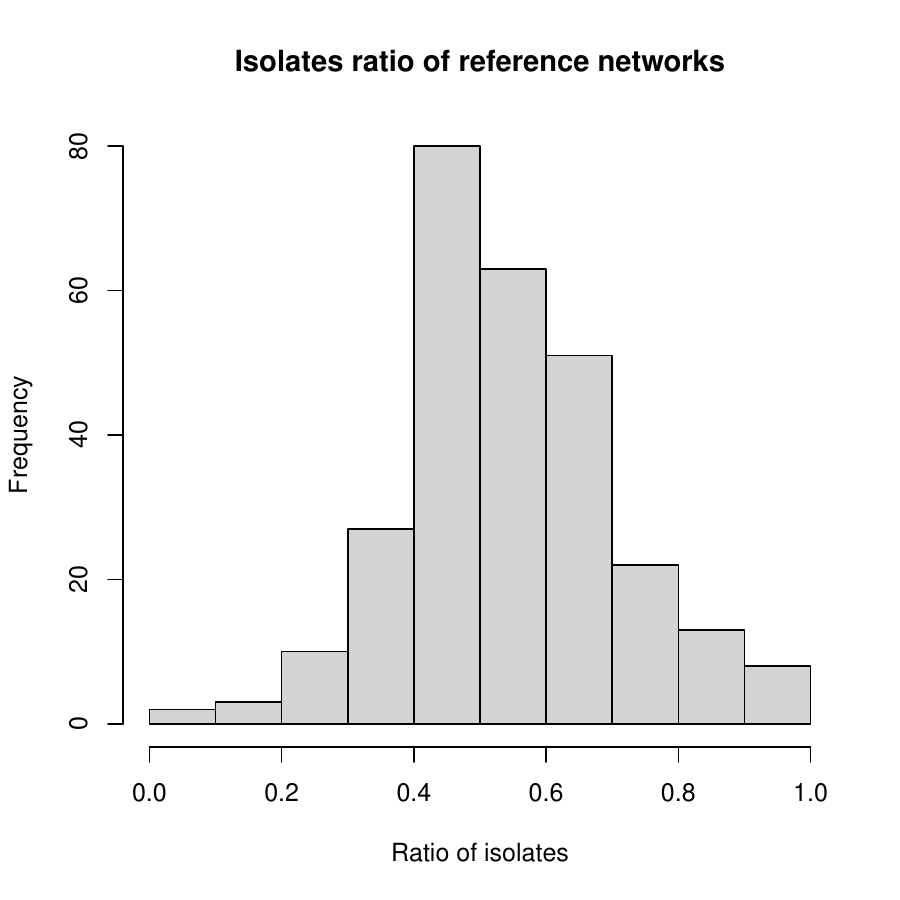}
  \caption{Ratio of isolates}
  \label{fig:net_iso}
\end{figure}

The first thing that should be noted is reference networks mostly have a number of isolates. Figure~\ref{fig:net_iso} shows the distribution of the ratio of isolates in 279 mention networks. It shows a nearly normal distribution of values with a mean above 0.5. This shows that almost half of the users in reference networks do not ever post a chat with a mention (`$@$'). The pattern is not that surprising as most users publish chat posts to comment about the ongoing streamings they are watching, which does not need to mention somebody. 

One notable observation is that reference networks typically include a significant number of isolates. Figure~\ref{fig:net_iso} illustrates the distribution of isolate ratios across 279 mention networks. The distribution follows a nearly normal pattern with a mean slightly above 0.5, indicating that nearly half of the users in these networks never post a chat with a mention (`$@$'). This pattern is understandable since many users contribute chat posts solely to comment on ongoing streams they are watching, without needing to mention others. However, in approximately 40 political streams, over 60\% of their interactions involve references to other viewers or streamers. This suggests highly engaged discussions in certain political streams, akin to the very intimate discussions often found in smaller Twitch streams \cite{mun2021str}. This also aligns with the earlier finding that most channels likely involve around 70 users who have posted at least one chat post (see Figure~\ref{fig:freq_user}).

\begin{figure}[h!]
    \centering
    \includegraphics[width=0.4\linewidth]{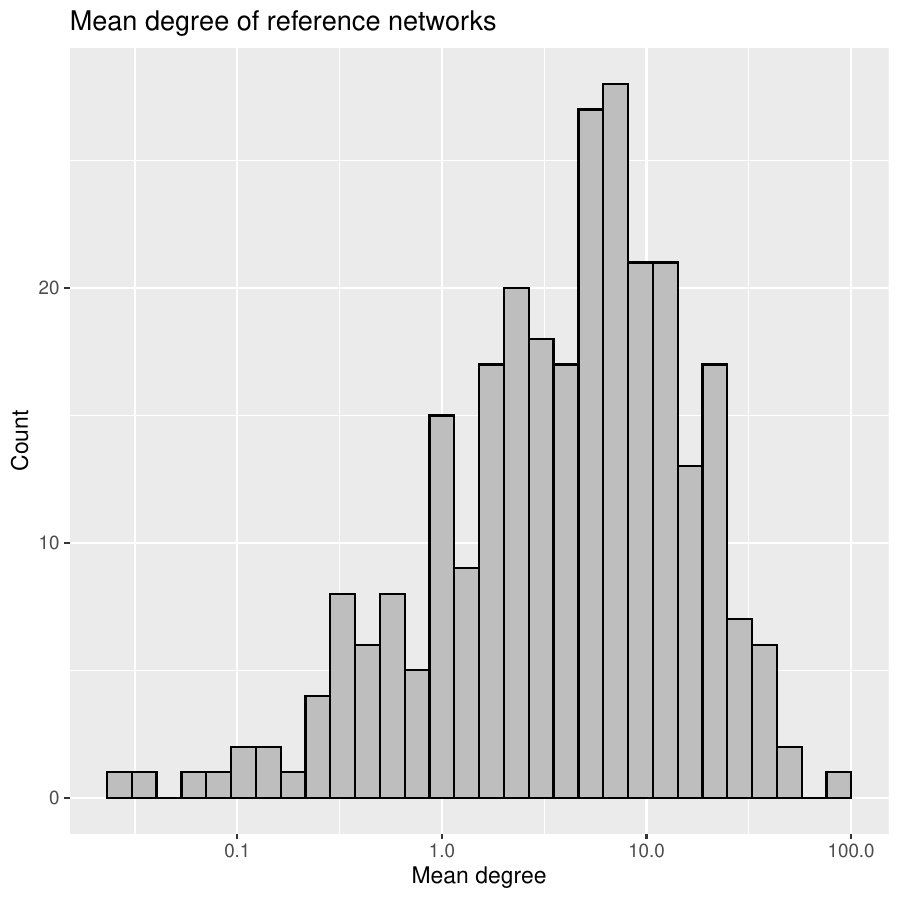}
  \caption{Distribution of mean node degrees. The x-axis is log transformed while labels show the raw values.}
  \label{fig:net_deg}
\end{figure}

\begin{figure}[h!]
    \centering
    \includegraphics[width=0.4\linewidth]{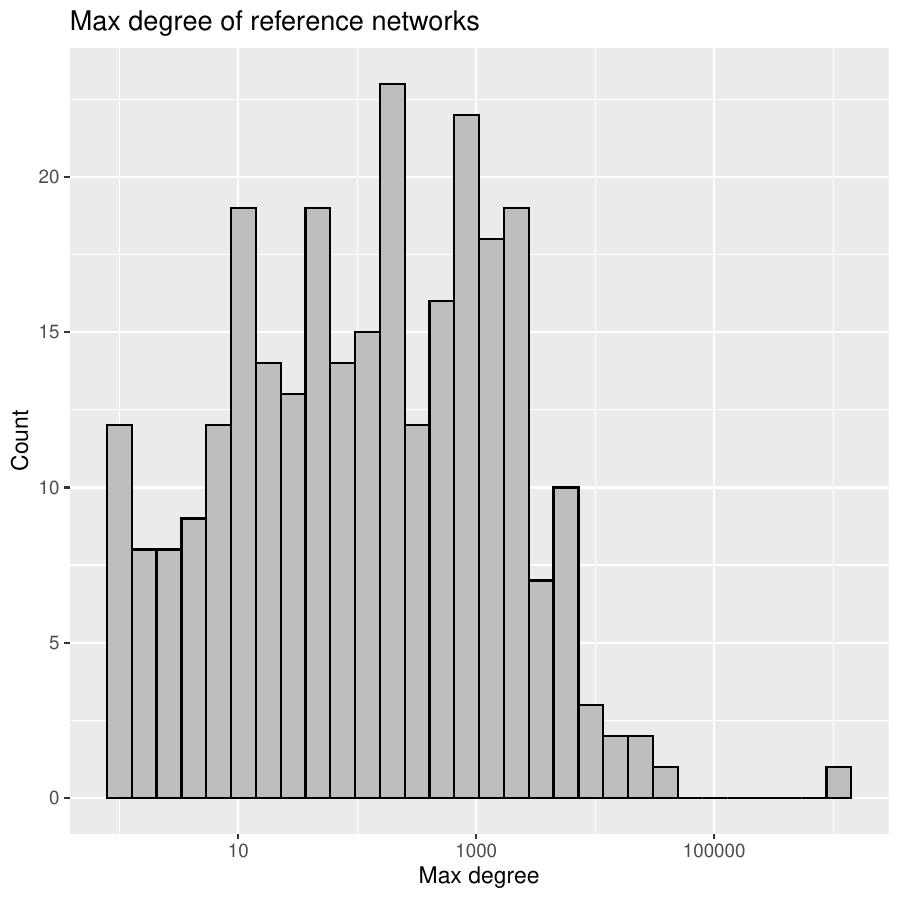}
    \includegraphics[width=0.4\linewidth]{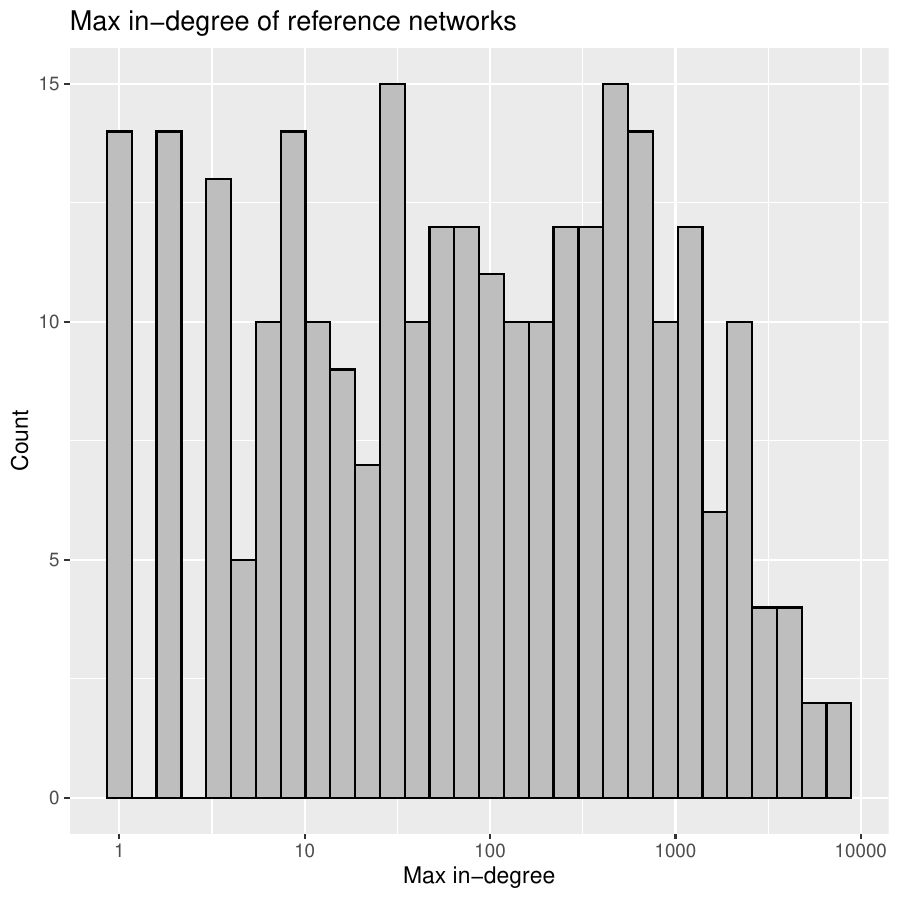}
    \includegraphics[width=0.4\linewidth]{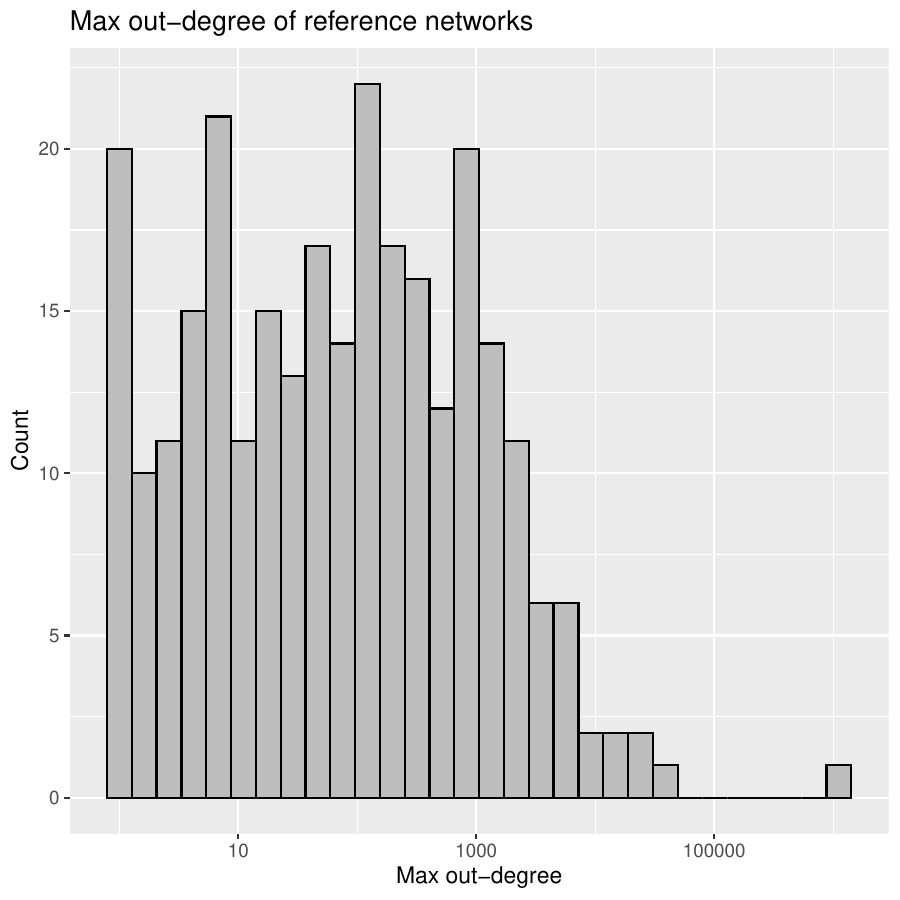}
  \caption{Distribution of max node degrees. The x-axis is log transformed while labels show the raw values.}
  \label{fig:net_degio}
\end{figure}

Then, how many chat posts mentioning other users do Twitch users send and receive? The histograms depicting the mean degree of all nodes in each mention network are presented in Figure~\ref{fig:net_deg}. The distribution of logged values indicates a nearly normal pattern, with a mean of 7.9 and the highest mean degree reaching 84.3. Figure~\ref{fig:net_degio} further illustrates distributions of maximum node degree, focusing separately on in-degree and out-degree. A notable observation is that the maximum out-degree values are generally higher than those of in-degree. This suggests that a few prolific users, who frequently mention others, significantly influence the overall degree distribution. Specifically, the peak values in the maximum all-degree histogram (upper left) align closely with those in the maximum out-degree histogram (bottom), indicating their substantial impact.

The mean value of the maximum all degree is 5130.5, with a median of 125, indicating significant skewness due to a few exceptionally active users, as evident from the histograms. These users, often channel owners or moderators, play roles such as answering questions during streams or enforcing streaming rules, contributing extensively to the network's edges. They can be considered opinion leaders within these reference networks, owing to their disproportionately large number of connections.

\begin{figure}[h!]
    \centering
    \includegraphics[width=0.4\linewidth]{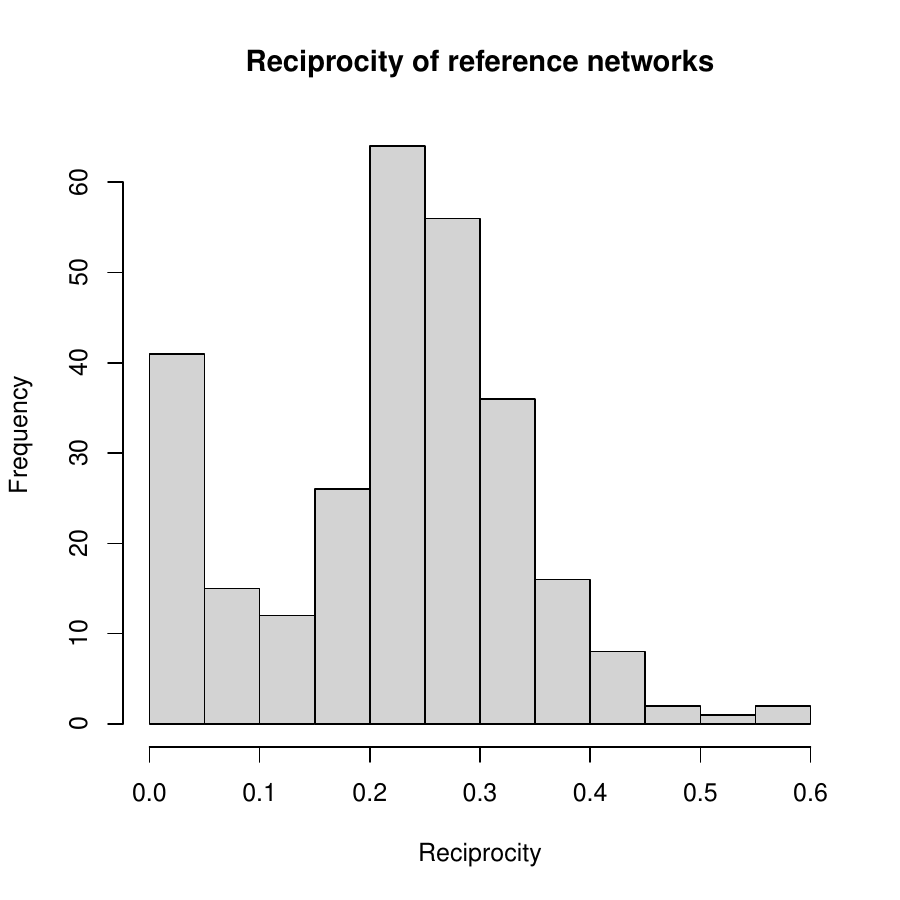}
  \caption{Distribution of reciprocity scores}
  \label{fig:net_rec}
\end{figure}
The reciprocity and modularity scores of all reference networks are depicted in Figure~\ref{fig:net_rec}. The reciprocity score represents the proportion of reciprocal ties within each network, while the modularity score indicates how connected nodes are within their communities. The mean reciprocity score is 0.22, indicating that, on average, 22\% of connections among nodes are reciprocal. This means approximately one-fifth of interactions within each reference network involve mutual mentions between users. This finding aligns with Figure~\ref{fig:net_degio}, which highlights the dominance of some highly active users in the communication network, where lower reciprocity is expected.

\begin{figure}[h!]
    \centering
    \includegraphics[width=0.4\linewidth]{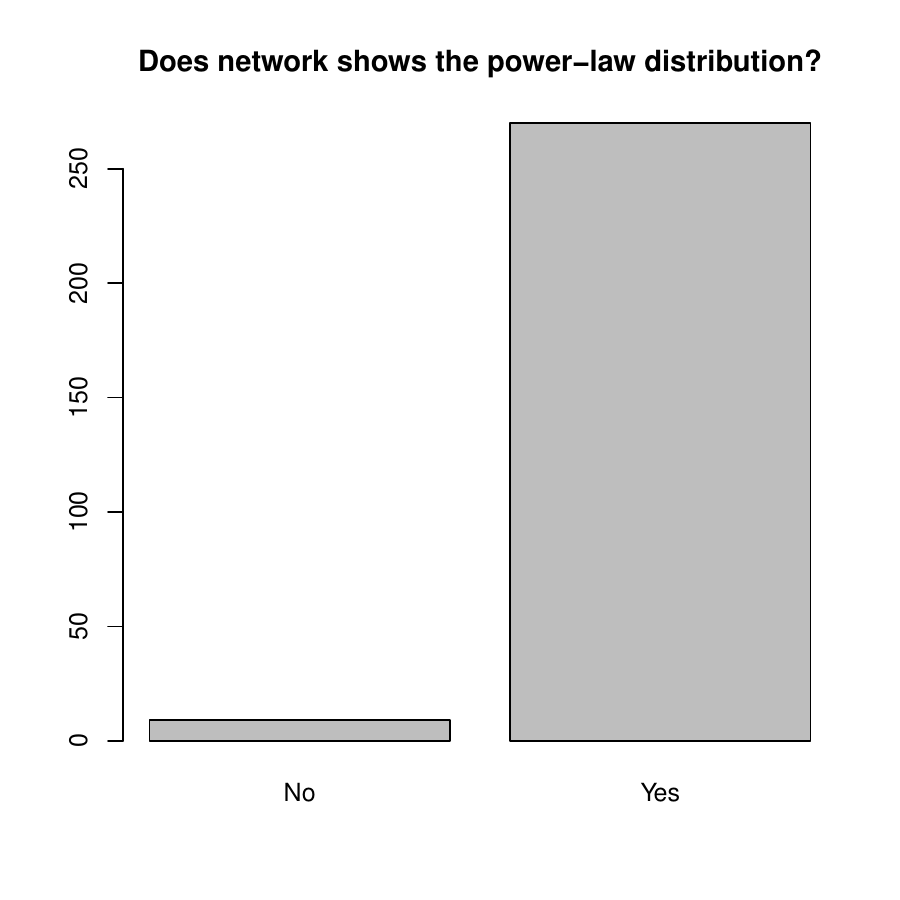}
  \caption{Does network shows the power-law degree distribution?}
  \label{fig:net_pow}
\end{figure}

Figure~\ref{fig:net_pow} demonstrates that many mention networks follow a power-law degree distribution. I fitted a power-law degree distribution to each network and used the Kolmogorov-Smirnov test to determine if we could reject the null hypothesis—that the network could be derived from this distribution. The results show that over 96\% of the networks conform to the fitted power-law degree distribution, similar to those observed on other social media platforms \cite{bodrunova2018power, martha2013study, myers2014information}. This implies that the structural characteristics of mention networks in political streaming chats on Twitch are not significantly different from those on other social media platforms. In other words, despite the technological differences inherent in real-time interaction and streaming chat, the fundamental structure of political communication networks on Twitch remains similar to those on other platforms.


\section{Conclusion}

In this article, I have addressed three main questions regarding Twitch politics: 
1) Who are the political Twitch streamers?
2) What content is covered in political streams?
3) How do audiences of political streams interact with each other? 
These questions were explored by focusing on the streaming chat function, which is central to the platform's communication technologies.

To answer the first question, I utilized the Twitch API and supervised machine-learning techniques to identify 574 political streamers. This study is the first comprehensive attempt to identify political actors on the platform, providing a valuable methodology for future research. The pipelines used to identify political streamers and collect chat data will be fully open and shared with this article.

For the second question, I employed topic modeling to examine the content of political streams. Since the content of political streams on Twitch was previously unknown, this study contributes by offering informative snapshots of this content. Notably, I found that identity-related topics are frequently discussed, highlighting how real-time interaction technology may influence topic choices among political actors. Future research could further explore identity-related topics, such as focusing on feminist streamers.

To address the third question, I created and analyzed user-reference networks within each political streamer's chatroom. The analysis revealed that a small number of audience members dominate the communication network by frequently referring to one another. Additionally, despite the technological differences of real-time interaction and streaming chat, the fundamental structure of political communication networks on Twitch resembles those on other platforms. Most user-reference networks follow a power-law distribution, similar to communication networks on other social media platforms \cite{bodrunova2018power, martha2013study, myers2014information}.

There are, however, areas for improvement. This study only analyzed text data from stream titles and profiles of streamers who used three specific game names. There may be political streamers who discuss political issues without explicitly indicating so in their titles or profiles. Addressing this limitation in future research could lead to a more thorough identification of political streamers and a deeper understanding of Twitch politics. Additionally, chat post data could be analyzed in various ways, such as focusing on toxicity. Further analyses could also explore the partisanship of political streamers and how user engagement in streaming chat varies based on the streamer's partisanship. These insights could contribute to the literature on both partisanship and political communication.

\section*{Supporting information}
\renewcommand{\thetable}{A\arabic{table}}
\renewcommand{\thefigure}{A\arabic{figure}}
\renewcommand{\thesubsection}{A\arabic{subsection}}
\setcounter{figure}{0}
\setcounter{table}{0}
\paragraph*{S1 Appendix A. Game name identification}
\label{S1}
The first step of the process is getting to know what ``game-names" political streamers use. ``Game-names" are tags of Twitch live-streaming that refer to what streamers are broadcasting. As Twitch is a mainly game-oriented platform, most of game-names are the names of games they are playing, such as League of Legends and Fortnite. However, there are game names that are not directly related to gaming activities, ranging from ``Just Chatting" to ``Cooking". \par
It is quite apparent that streamers who stream with the game name ``Politics" can be classified as political streamers and can be added to the list. In addition, based on the information from the site ``Twitchmetrics", which provides various information of Twitch streamers including rankings by category based on the information they received through Twitch API, I was able to know that political streamers also use the game name ``Talk Shows and Podcasts" \cite{twitchmetrics2022}. However, there are also some streamers who stream political content without using ``Politics" or ``Talk Shows and Podcasts" as the game name of their streams. The best example would be Hasanabi, who has been thoroughly described in the previous section: He never broadcast with ``Politics". Instead, he uses ``Just Chatting" as the game name when he is streaming about political issues. This might indicate the possibility that I might be able to find political streamers out of all Twitch streamers who stream with ``Just chatting" tag. The challenging part is that most of the streamers who do not explicitly stream their gaming activity (sometimes they even stream their gaming activity with ``Just Chatting" game name) but do something else choose to use "Just Chatting" as their game name. The topics vary from talk shows by herself or with other streamers to vlogging their traveling activities. In other words, when I use Twitch API to get information on live streaming with ``Just Chatting" game name, I will get a list of streamers who stream very different topics. But, before including ``Just Chatting" game name to the list of game names that might have political streamers, I conducted preliminary analysis on ``Just Chatting" streamers to see whether there are political streamers who are using the game name.\par
Using Twitch API ``Get Streams" function, I have retrieved the list of streamers who stream in English with the game name ``Just Chatting".\footnote{The list has been retrieved at 12:00 PM, June 30th, 2021} As there are a substantial number of streamers who have only a single viewer, I stopped when the list ends up with a streamer with a single viewer count. I was able to identify 873 streamers who were streaming at the time by making ``Get Stream" requests through Twitch API. ``Get Stream" request retrieves various information about the ongoing streaming: user id of the streamer, viewer count, the title of the streaming, URL of a thumbnail image, etc. Then, how can I know whether streaming covers political content or not? I found thumbnail images and titles of streamings can be helpful.\par
By using URL information in the retrieved data of live streaming, I was able to download all thumbnails of 873 streamings. Thumbnails are basically the screen viewers were watching at the time I made the ``Get Stream" request. Therefore, by downloading thumbnail images, I can retrieve visual information of ongoing streamings. By combining visual information and text information from streaming titles, I was able to identify some political streamings.  
\begin{figure}[h!]
    \centering
    \includegraphics[scale=0.3]{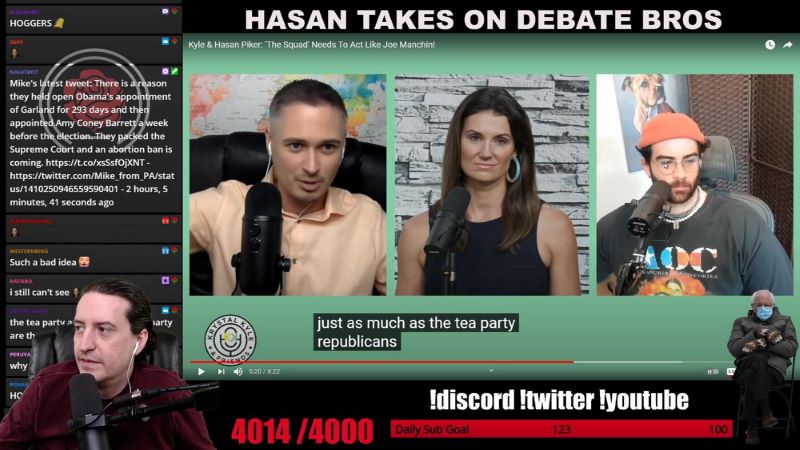}
    \includegraphics[scale=0.3]{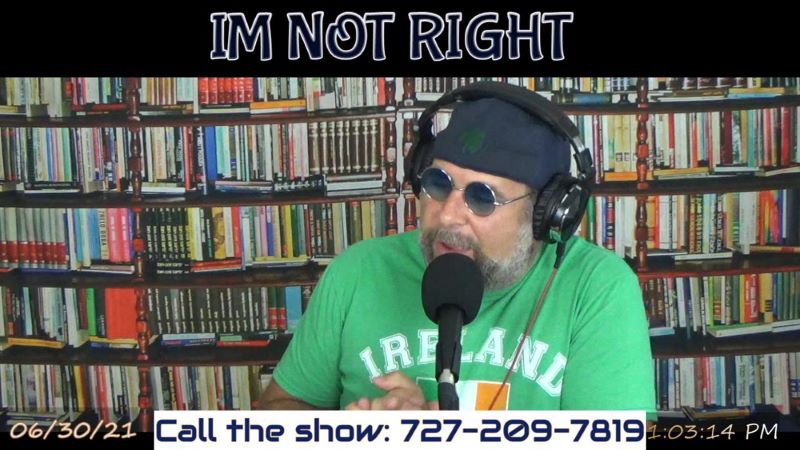}
    \includegraphics[scale=0.3]{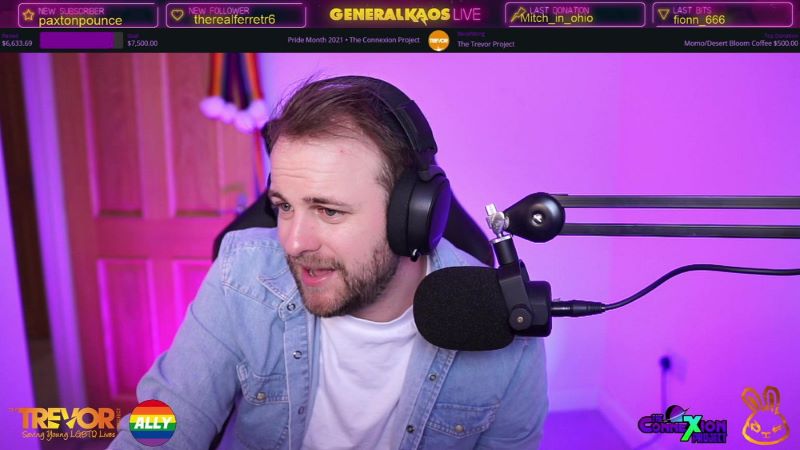}
    \includegraphics[scale=0.3]{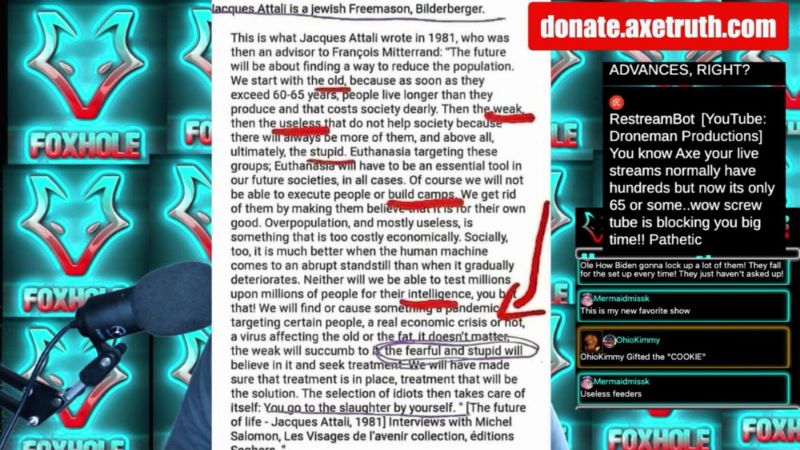}
    \caption{Example of political streamings with "Just Chatting" game name}
    \end{figure}
Figure A1 shows the thumbnail images of political streamings. Upper left of figure 1 is a thumbnail image of the streaming by ``Central\_Committee" with the title ``DNC WORKING TO BLOCK NINA TURNER | HASAN MALDING OVER DEBATE BROS | NYC MAYOR'S RACE VOTE COUNT MASSIVE ERROR | ".  Upper right is a thumbnail image of the streaming by ``ImNotRightNetwork" with the title ``Where are our freedoms?". Lower left is a thumbnail image of the streaming by ``GeneralKaosLive" with the title of ``(rainbow emote) TREVOR PROJECT FUNDRAISER (rainbow emote)Helping to prevent suicide and giving back to the LGBTQ+ communty - !Charity !CXP ". The lower right is a thumbnail image of the streaming by ``Axe\_Truth" with the title of ``Discouraged \& Division in the ranks Candace Owens vs Kim Klacik". When we use both textual information from the titles and visual information from the thumbnail images, they can definitely be classified as political streamers. All of the titles touch on some political terms, such as ``Mayor's Race", ``freedom", ``LGBTQ+", ``Candace Owens vs Kim Klacik". Also, the composition of thumbnail images shares some characters. The upper left and lower right both show political content (one is visual and the other is textual). And except for the lower right, all three streamers use a similar broadcasting setting: a very huge microphone with their face exposed. Overall, there are political streamers who broadcast with the game-name ``Just Chatting".\par

\renewcommand{\thetable}{B\arabic{table}}
\renewcommand{\thefigure}{B\arabic{figure}}
\renewcommand{\thesubsection}{B\arabic{subsection}}
\setcounter{figure}{0}
\setcounter{table}{0}
\paragraph*{S2 Appendix B. Supervised machine learning classifier for identifying political streamers}
\label{S2}
\begin{table}[hbtp!]
\centering
\caption{ML Models Performance}
\begin{tabular}{|l|l|l|l|l}
\cline{1-4}
Model                                                                                   & Precision & Recall & F-1   &  \\ \cline{1-4}
Logit + Count                                                                           & 0.822     & 0.721  & 0.760 &  \\ \cline{1-4}
Logit + TFIDF                                                                           & 0.907     & 0.640  & 0.703 &  \\ \cline{1-4}
\begin{tabular}[c]{@{}l@{}}XGBoost + Count\\    \\ (learning rate =   0.5)\end{tabular} & 0.855     & 0.745  & 0.788 &  \\ \cline{1-4}
\begin{tabular}[c]{@{}l@{}}XGBoost + TFIDF\\    \\ (learning rate =   0.5)\end{tabular} & 0.868     & 0.732  & 0.781 &  \\ \cline{1-4}
\end{tabular}
\end{table}
\begin{figure}[h!]
    \centering
    \includegraphics[width=1.0\linewidth]{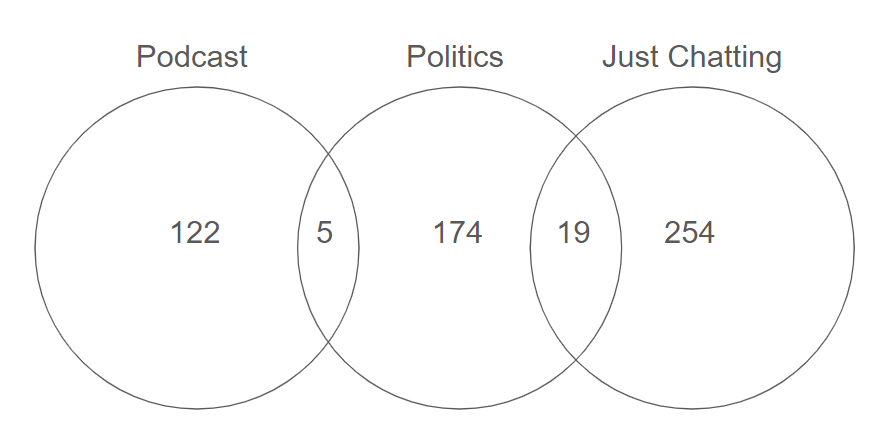}
    \caption{Venn Diagram of Political Streamers}
\end{figure}
I have trained logistic regression and XGBoost classifiers using the train data. And I have specified models with the count and TF-IDF vectorizer for each algorithm. As the target value of the dataset is infrequent, I have used precision, recall, and F-1 score, which are known to be more appropriate to evaluate imbalanced data, to evaluate the performance of classifiers. Table A1 shows the performance of the 5-fold cross-validation results of each model. XGboost models mostly outperform logistic regression models in all three criteria, while it is hard to say which vectorizer is outperforming. Therefore, I have four hyperparameters of the XGBoost models with different vectorizers: \textit{colsample\_bytree, learning\_rate, max\_depth} and \textit{n\_estimators}. I have used the Gridsearch method to find a combination of hyperparameters to produce the best macro F1 score. The best score of each model with different vectorizers looks almost identical: 0.796 for the model with TFIDF vectorizer and 0.795 for the model with count vectorizer. Although the difference is negligible, I have used the model with TFIDF vectorizer for machine labeling the rest of the unlabeled data as it shows slightly better performance. Through this process, I was able to identify additional 48 political accounts from 45,477 just chatting streamer accounts. Adding 550 political accounts identified during the hand-coding process, I was able to identify 598 political accounts in total.\par
However, there can be duplicates in the list, as the streamers are able to switch the `game name' of their streaming whenever they want. For example, a hypothetical streamer A can stream one day with ``Just Chatting" game name on the other day with ``Politics" game name. In this case, streamer A would be in the intersection between the ``Politics" set and ``Just Chatting" set in Figure A2. 
I was able to find 24 streamers who have ever streamed with more than a single game name. There are 19 accounts that have at least once streamed both in ``Just Chatting” and ``Politics” and 5 streamers who have ever streamed with ``Talk Shows and Podcast” and ``Politics". After dealing with the duplicates, I was finally able to identify 574 unique political accounts on Twitch. Figure 3 shows the breakdown of these 574 political streamers: 254 streamers have streamed only with ``Just Chatting", 174 streamers have streamed only with ``Politics", 122 streamers have streamed only with ``Talk Shows and Podcasts", 19 streamers have streamed with both ``Just Chatting" and ``Politics", and 5 streamers have streamed with both ``Talk Shows and Podcast" and ``Politics".

\newpage
\renewcommand{\thetable}{C\arabic{table}}
\renewcommand{\thefigure}{C\arabic{figure}}
\renewcommand{\thesubsection}{C\arabic{subsection}}
\setcounter{figure}{0}
\setcounter{table}{0}
\paragraph*{S3 Appendix C. Extensive lists of political topics}
\label{S3}

\leavevmode\vspace{1em}

\begin{figure}[htbp]
    \centering
    \includegraphics[width=0.7\linewidth]{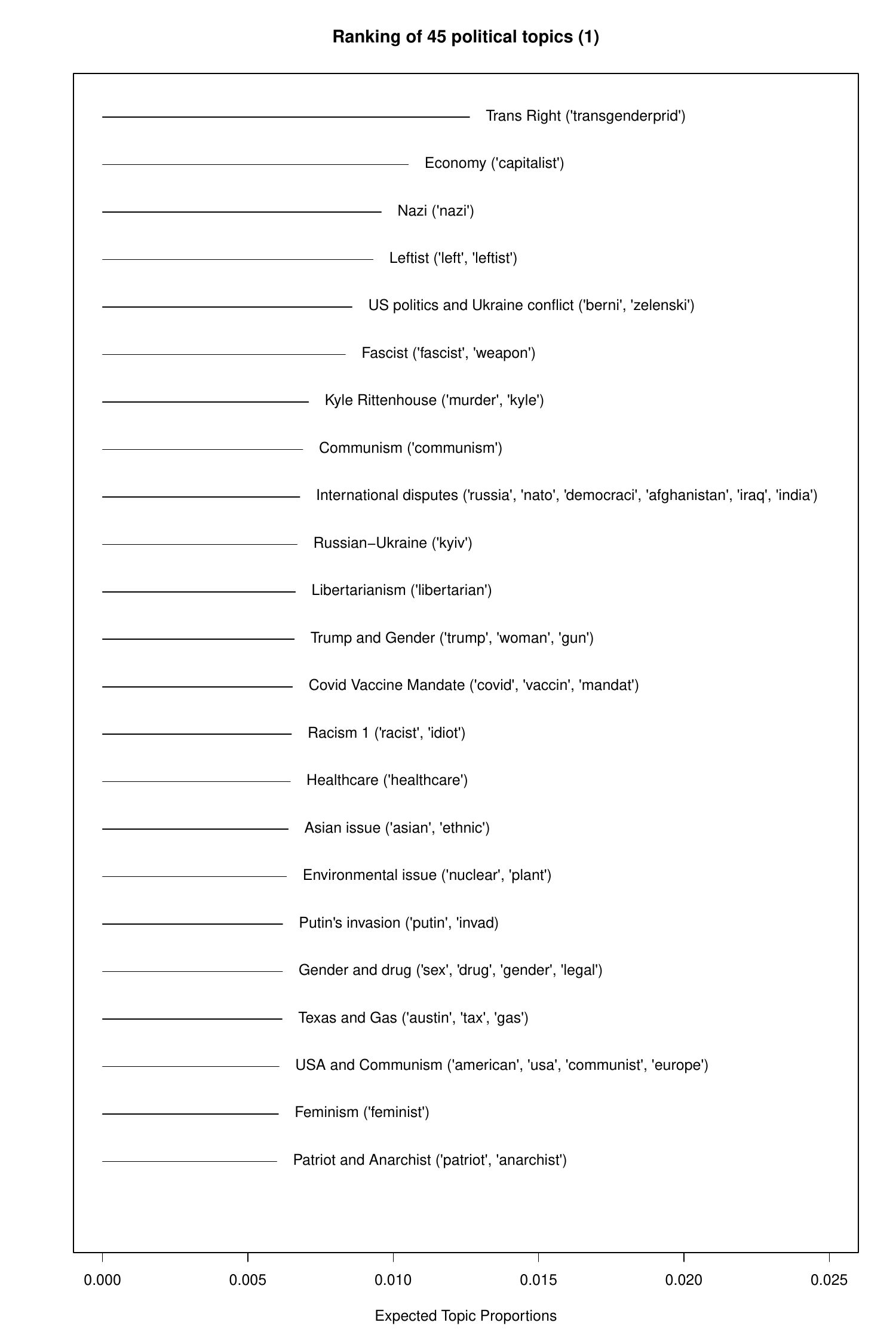}
    \caption{Identified Political Topics (1)}
\end{figure}
\begin{figure}[h!]
    \centering
    \includegraphics[width=0.7\linewidth]{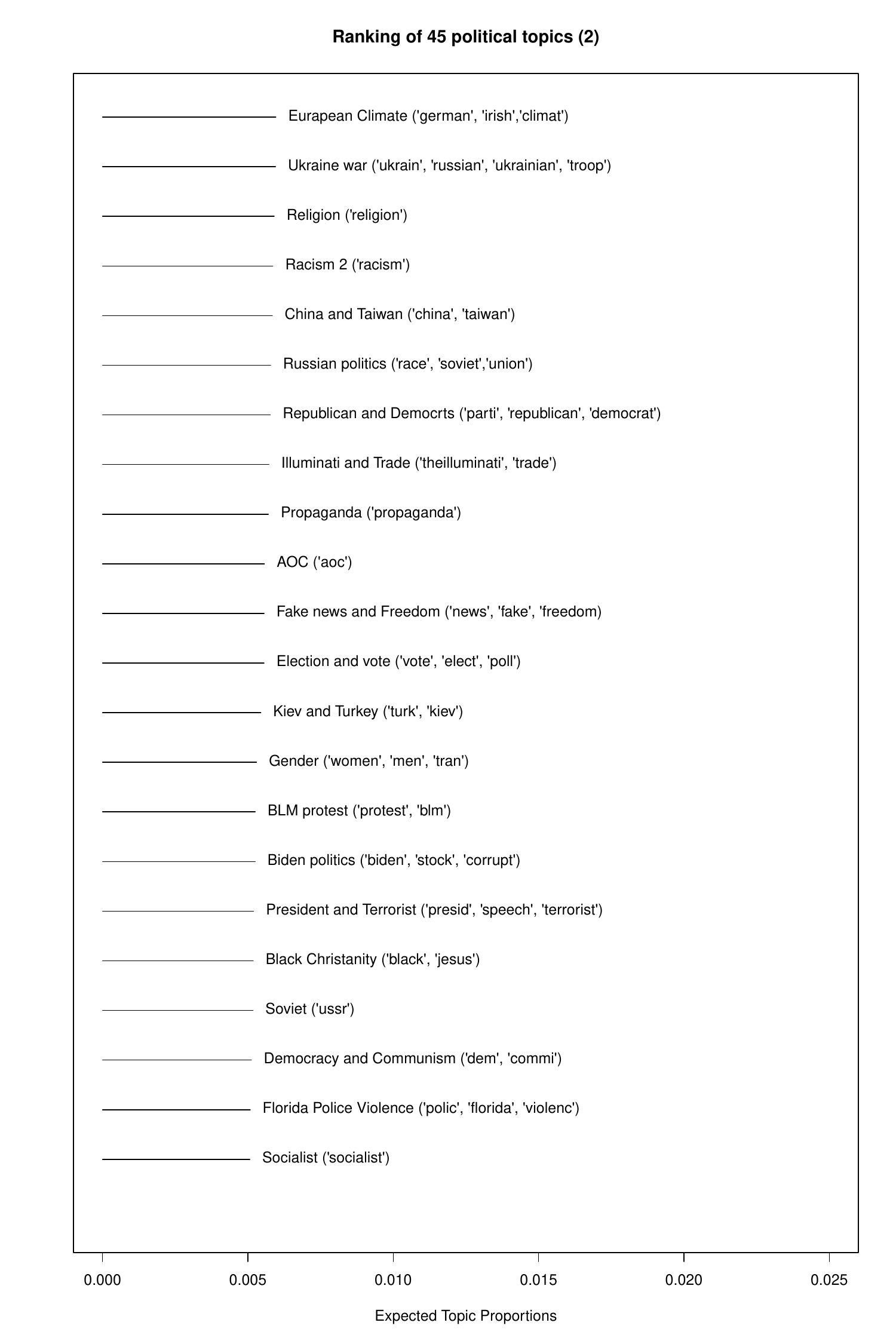}
    \caption{Identified Political Topics (2)}
\end{figure}

%
%
%
\newpage
\section*{Acknowledgments}
Thanks to one of the authors of Flores et al. (2019), Joseph Seering, it was possible to successfully go through the data collection process. Streaming chat posts collection was largely done based on the codes in his Github:\url{https://github.com/josephseering/twitchtext}
\nolinenumbers
\bibliography{name}

\end{document}